\documentclass[11pt,a4paper]{scrartcl}
\usepackage[round]{natbib}

\usepackage[top=3cm, bottom=3cm, left=3cm, right=3cm]{geometry}

\usepackage{booktabs}

\usepackage{color}
\usepackage{latexsym}              
\usepackage{amsmath}               
\usepackage{amssymb}               
\usepackage{amsfonts}              
\usepackage{amsthm}                
\usepackage{multirow}
\usepackage{tikz}                  
\usetikzlibrary{arrows,positioning,shapes}
\usepackage{tcolorbox}
\usepackage{placeins}
\usepackage{subcaption}
\usepackage{algorithm}
\usepackage{bm}

\usepackage[english]{babel} 

\RequirePackage[%
  pdfstartview=FitH,%
  breaklinks=true,%
  bookmarks=true,%
  colorlinks=true,%
  linkcolor= blue,
  anchorcolor=blue,%
  citecolor=orange,
  filecolor=blue,%
  menucolor=blue,%
  urlcolor=blue%
  ]{hyperref}
  \usepackage[capitalize,nameinlink,noabbrev]{cleveref}

\renewenvironment{proof}[1][\proofname]{{\bfseries #1.}}{\qed \\ }

\makeatother

\theoremstyle{plain}  
\newtheorem{theorem}{Theorem}[section]
\newtheorem{definition}[theorem]{Definition}

\newtheorem{lemma}[theorem]{Lemma}
\newtheorem{proposition}[theorem]{Proposition}

\newtheorem{corollary}[theorem]{Corollary}
\newtheorem{remark}[theorem]{Remark}

\newtheorem{assumption}[theorem]{Assumption}

\newcommand{\bbE}{\mathbb{E}}

\newcommand{\bbR}{\mathbb{R}}

\newcommand{\Real}{{\mathbb R}}

\DeclareMathOperator*{\supp}{supp}
\DeclareMathOperator{\vol}{vol}

\newcommand{\fn}[3]{f_{#1,#2}(#3)}
\newcommand{\fntilde}[2]{\widetilde{f}_{#1}(#2)}
\newcommand{\fntildebis}[2]{\widetilde{f}_{#1}(#2)}

\newcommand{\fintegral}[2]{m_{#1}(#2)}
\newcommand{\ftildeintegral}[2]{\widetilde m_{#1}(#2)}

\newcommand{\1}[1]{\mathbf{1}_{\{#1\}}}

\newcommand{\yn}{\overline{y}_{n}}
\newcommand{\Yn}{\overline{Y}_{n}}

\def\[#1\]{\begin{align}#1\end{align}}

\newcommand{\iid}{i.i.d.\ }

\newcommand{\KL}{\text{\texttt{KL}}}

\newcommand{\dHaus}{d_{\mathrm{H}}}

\newcommand{\CVy}{C_{\alpha,n}^{\mathrm{V}}(y_{1:n})}
\newcommand{\CeVy}{C_{\alpha,n}^{\mathrm{eV}}(y_{1:n})}
\newcommand{\CeVYtheta}[1]{C_{\alpha,n}^{\mathrm{eV}}(Y^{(#1)}_{1:n})}
\newcommand{\CPy}{C_{\alpha,n}^{\mathrm{P}}(y_{1:n})}
\newcommand{\CVY}{C_{\alpha,n}^{\mathrm{V}}(Y_{1:n})}
\newcommand{\CeVY}{C_{\alpha,n}^{\mathrm{eV}}(Y_{1:n})}

\newcommand{\CV}{C_{\alpha,n}^{\mathrm{V}}}
\newcommand{\CeV}{C_{\alpha,n}^{\mathrm{eV}}}

\newcommand{\thetatrue}{\theta_\ast}

\usepackage[inline]{enumitem}
\newlist{enuminline}{enumerate*}{1}
\setlist[enuminline]{label=\textbf{\arabic*})}

\begin{document}
\title{Confidence sequences with informative, bounded-influence priors}

\author{
  Stefano Cortinovis\textsuperscript{1,*} \quad
  Valentin Kilian\textsuperscript{1,*} \quad
  François Caron\textsuperscript{1} \\
  \small
  \textsuperscript{1}Department of Statistics, University of Oxford \\
    \small
  \textsuperscript{*}Equal contribution. Order decided by coin toss. \\
    \small
\texttt{cortinovis@stats.ox.ac.uk}, \quad
  \texttt{kilian@stats.ox.ac.uk}, \quad
  \texttt{caron@stats.ox.ac.uk}
}
\date{}

\maketitle

\begin{abstract}
Confidence sequences are collections of confidence regions that simultaneously cover the true parameter for every sample size at a prescribed confidence level.
Tightening these sequences is of practical interest and can be achieved by incorporating prior information through the method of mixture martingales.
However, confidence sequences built from informative priors are vulnerable to misspecification and may become vacuous when the prior is poorly chosen.
We study this trade-off for Gaussian observations with known variance.
By combining the method of mixtures under a prior with polynomial or exponential tails with the extended Ville's inequality, we construct confidence sequences that are sharper than their non-informative counterparts whenever the prior is well specified, yet remain bounded under arbitrary misspecification.
Along the way, we provide general sufficient conditions under which extended Ville confidence sequences are convex sets, and show that the Bayesian posterior mean always lies inside the confidence regions obtained by the method of mixtures, thereby providing a natural accompanying point estimator.
The theory is illustrated through simulations with several classical priors.
\end{abstract}


\section{Introduction}

A confidence sequence (CS) is a sequence of confidence regions that is simultaneously valid over all sample sizes.
Although the idea has a long history dating back to the work of \citet{Darling1967}, \citet{Robbins1970} and \citet{Lai1976}, confidence sequences have attracted renewed and growing interest in recent years.
This revival is driven by applications in sequential experimentation and decision-making such as group-sequential clinical trials, A/B testing, and bandit problems \citep{Jennison1984,Johari2022,Jamieson2018}.
Methodologically, recent work has extended time-uniform inference beyond the classical mean problem to nonparametric settings, quantiles, multinomial models, asymptotic regimes, and general $M$-estimators \citep{Howard2021,Howard2022,Lindon2022,WaudbySmith2024,Schreuder2020}; see \citet{Ramdas2023} for an overview.

A standard route to constructing a CS is to invert a family of sequential tests obtained by thresholding nonnegative supermartingales, or more generally extended nonnegative supermartingales.
Coverage then follows from Ville's inequality \citep{Ville1939} in the former case, and from the extended Ville's inequality \citep{Wang2023} in the latter;
we refer to the resulting procedures as Ville confidence sequences (VCSs) and extended Ville confidence sequences (eVCSs), respectively.
A versatile way to build the required processes is the classical method of mixtures, introduced by \citet{Ville1939} and further developed by \citet{Wald1945}, \citet{Darling1967}, \citet{Robbins1970b}, \citet{Robbins1970} and \citet{Lai1976}; see \citet{Kaufmann2021} for a modern treatment.
In this approach, likelihood ratios, or more general test supermartingales, are averaged with respect to a mixing measure over the parameter space.
This construction underlies a range of modern CS methods, with applications including sampling without replacement \citep{WaudbySmith2020}, Gaussian processes \citep{Neiswanger2021}, multinomial data \citep{Lindon2022}, and prediction-powered inference \citep{Kilian2025}.

\begin{figure}[ht]
  \centering
  \includegraphics[width=\textwidth]{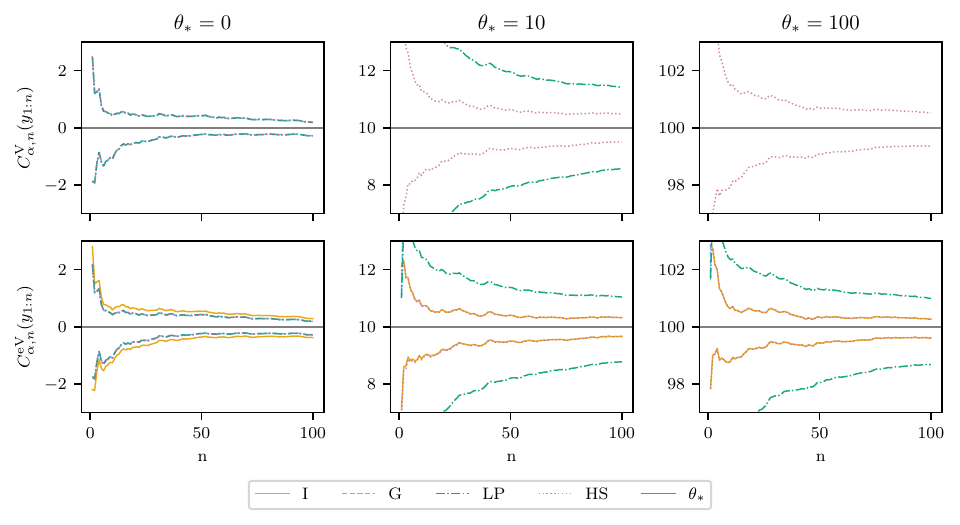}
  \caption{
    Comparison of $90\%$ VCSs (top row) and eVCSs (bottom row) resulting from a single realisation of $Y_{1:n}$ from \eqref{eq:dgp} with $\thetatrue \in \{0, 10, 100\}$ and $\sigma^2 = 1$, under Gaussian (G), Laplace (LP) and horseshoe (HS) priors with location $0$ and scale $0.1$, as well as the improper prior (I) discussed in \cref{sec:Gaussianimproper} (only extended Ville).
    Larger values of $|\thetatrue|$ correspond to stronger prior--data conflict.
    As $\thetatrue$ grows, VCSs diverge regardless of the prior (\cref{rk:Villepart}), whereas eVCSs remain bounded for priors with exponential (LP) or polynomial (HS) tails (\cref{thm:main}).
    When $\thetatrue = 100$, both CSs under the G prior exceed the panel bounds (rightmost column), while the eVCS under the HS prior is indistinguishable from the improper prior's (bottom right plot).
    \cref{fig:cs_intro_long} in \cref{sec:additional_results} replicates this figure with $n$ up to $1000$.
  }
  \label{fig:cs_intro}
\end{figure}

The choice of mixing measure has a major effect on the efficiency of the resulting CS.
When it places substantial mass near the true parameter, the resulting regions can be markedly smaller than those obtained from diffuse or non-informative choices.
In this sense, the mixing measure plays the role of a prior, encoding problem-specific information into the construction;
we use the two terms interchangeably in the remainder of the paper.
At the same time, although coverage is guaranteed for any fixed mixing measure, severe prior--data conflict can make the resulting regions impractically large under standard constructions, as illustrated in \cref{fig:cs_intro}.
Our objective is to retain the efficiency gains of informative priors while guarding against such undesirable behaviour.

Concretely, let $(Y_i)_{i\ge1}$ be independent and identically distributed (i.i.d.) Gaussian random variables,
\begin{equation}
  Y_i \sim \mathcal{N}\bigl(\thetatrue,\sigma^2\bigr), \qquad i\ge1,
  \label{eq:dgp}
\end{equation}
with unknown mean $\thetatrue$ and known variance $\sigma^2$.
We seek a CS $(C_{\alpha,n}(y_{1:n}))_{n\ge1}$ for $\thetatrue$ satisfying
\begin{align}
  \Pr\bigl\{\thetatrue \in C_{\alpha,n}(Y_{1:n}) \text{ for all } n\ge1\bigr\} \ge 1-\alpha,
  \label{eq:coverage}
\end{align}
and meeting two desiderata:
\begin{itemize}
  \item \textbf{Efficiency under prior--data agreement.}
  If the prior places substantial mass near the true parameter, the resulting regions are tighter than those produced by diffuse or non-informative choices.
  \item \textbf{Robustness to prior--data conflict.}
  For each fixed $n$, the volume of $C_{\alpha,n}(y_{1:n})$\footnote{For a given $n$, the volume of $C_{\alpha,n}(y_{1:n})$, denoted by $\vol(C_{\alpha,n}(y_{1:n}))$, is its Lebesgue measure. In the one-dimensional case, this corresponds to the length of the region.} remains uniformly bounded over all data vectors $y_{1:n}\in\mathbb R^n$.
\end{itemize}

Our main contribution (\cref{thm:main}) shows that these two goals can be simultaneously achieved by combining
\begin{enuminline}
  \item the method of mixtures under a prior with exponential or polynomial tails and
  \item the extended Ville's inequality of \citet{Wang2023}.
\end{enuminline}
In particular, for priors with polynomial tails, each confidence region converges to that obtained from a constant improper prior as the degree of prior--data conflict grows, providing a procedure that smoothly reverts to a non-informative baseline in the worst case.
By contrast, we show that Ville's original inequality cannot deliver such robustness: under \emph{any} proper prior, the volume of the resulting VCS becomes arbitrarily large in the presence of conflict (\cref{rk:Villepart}).
\Cref{fig:cs_intro} provides an illustrative summary of these results.

These results connect anytime-valid inference with the classical literature on Bayesian robustness \citep{DeFinetti1961,Lindley1968,Strawderman1971,Berger1980}, and in particular with the theory of bounded-influence priors for Bayesian posteriors and credible intervals \citep{Dawid1973,Pericchi1992,Pericchi1995}, where polynomial- and exponential-tailed priors are known to play an analogous role.

We also provide two further developments on eVCSs of independent interest: general sufficient conditions on the likelihood and prior under which eVCSs are convex sets (\cref{th:convexityeVCS}), and a one-parameter family of improper-prior eVCSs indexed by a tail parameter $\kappa$, which generalises the construction of \citet{Wang2023} and arises as the limiting object of the theorem.
Separately, we show that the Bayesian posterior mean always lies inside the eVCS, and thus inside the VCS whenever the prior is proper, providing a natural point estimator to accompany these confidence sequences (\cref{prop:BAestimatorinCS}).

The remainder of the paper is organised as follows.
\Cref{sec:ville} provides background on Ville's inequality and its extended form, and reviews how these inequalities combine with the method of mixtures to construct confidence sequences.
For general likelihood and prior, we provide simple sufficient conditions under which eVCSs are convex sets (\cref{th:convexityeVCS});
\Cref{sec:Gaussian} specialises the mixture construction to the scalar Gaussian mean model \eqref{eq:dgp}, derives closed-form VCS and eVCS expressions under a Gaussian prior, and introduces the one-parameter family of improper-prior eVCSs (\cref{sec:Gaussianimproper}).
Our main contribution appears in \cref{sec:robustCS}, where we construct eVCSs from informative, bounded-influence priors and establish their robustness to prior--data conflict.
\Cref{sec:estimator} addresses the choice of point estimator to report alongside a confidence sequence, and shows that the Bayesian posterior mean always lies in the corresponding eVCS, and hence also in the VCS whenever the prior is proper.
Simulations illustrating the theory are presented in \cref{sec:exp}, and \cref{sec:discussion} concludes.
Proofs of secondary results and additional background are deferred to the Appendix.

\paragraph{Notations.}
For two real-valued functions $f$ and $g$ defined on $\bbR$, we write $f(x)\sim g(x)$ as $x\to\infty$ for $\lim_{x\to\infty} f(x)/g(x)=1$.
For a subset $C$ of $\bbR$ and $y\in\bbR$, let $C+y=\{x+y \mid x \in C\}$.
For two closed subsets $C_1$ and $C_2$ of $\bbR$, let $\dHaus$ denote the Hausdorff distance, $\dHaus(C_1,C_2)=\max\{\sup_{x\in C_1}\inf_{y\in C_2} |x-y|, \sup_{y\in C_2}\inf_{x\in C_1} |x-y| \}$.
For a collection of closed subsets $(C_1(y))_{y\in\bbR}$ and a closed subset $C_2$ of $\bbR$, we write $\lim_{y\to\infty}C_1(y)= C_2$ for $\lim_{y\to\infty}\dHaus(C_1(y),C_2)= 0$.
When $C_2=[a,b]$ is a closed interval, with $a<b$, $\lim_{y\to\infty} C_1(y)= C_2$ iff, for all $\epsilon\in(0,\frac{b-a}{2})$, there exists $y_0$ such that $[a+\epsilon,b-\epsilon]\subseteq C_1(y) \subseteq [a-\epsilon,b+\epsilon]$ for all $y>y_0$. 

\section{General mixture constructions and convexity of eVCSs}
\label{sec:ville}

This section collects the technical background used throughout the paper.
\Cref{sec:Villepart,sec:confidencesequenceglobal} review standard material: the (extended) Ville's inequality and the method-of-mixtures construction of Ville and extended Ville confidence sequences.
\Cref{sec:convexity} then presents our first new result, providing simple sufficient conditions under which eVCSs are convex sets, and in particular intervals in the scalar case.

\subsection{(Extended) supermartingales and (extended) Ville's inequality}
\label{sec:Villepart}

The confidence sequences considered in this paper are all built on Ville's inequality or its recent extension by \citet{Wang2023}.
Both may be viewed as anytime-valid analogues of Markov's inequality for nonnegative supermartingales.
We first recall the classical setting.
\begin{definition}[Nonnegative supermartingale]
    An adapted process $(X_{n})_{n\geq1}$ is a nonnegative supermartingale with respect to the filtration $(\mathcal{F}_{n})_{n\geq1}$ if $X_{n}\geq0$ a.s., $\mathbb{E}[X_{1}]<\infty$ and
    \begin{equation*}
        \mathbb{E}[X_{n+1} \mid \mathcal{F}_{n}] \leq X_{n}\text{ a.s. for all }n\geq 1.
    \end{equation*}
    If equality holds for all $n\geq 1$, then $(X_{n})_{n\geq1}$ is a nonnegative martingale.
\end{definition}
\begin{proposition}[Ville's inequality; \citealp{Ville1939}]
    \label{thm:VilleCorollary}
    Let $(X_{n})_{n\geq1}$ be a nonnegative supermartingale.
    For any $\alpha\in(0,1]$,
    \begin{equation*}
        \Pr\left(  X_{n}\leq\frac{\mathbb{E}[X_{1}]}{\alpha}\text{ for all } n\geq1\right) \geq 1-\alpha.
    \end{equation*}
\end{proposition}

\citet{Wang2023} presented a generalisation of Ville's inequality that yields slightly tighter bounds for nonnegative supermartingales and still applies when $\mathbb{E}[X_{1}]=\infty$ and/or $X_1=\infty$ with positive probability.
\begin{definition}[Extended nonnegative supermartingale; {\citealp[Definition 3.1]{Wang2023}}]
    An adapted process $(X_{n})_{n\geq1}$ is an extended nonnegative supermartingale with respect to the filtration $(\mathcal{F}_{n})_{n\geq1}$ if $X_{n}\in [0,\infty)\cup\{\infty\}$ a.s.~for all $n\geq1$, and
    \begin{equation*}
        \mathbb{E}[X_{n+1}\mid\mathcal{F}_{n}]\leq X_{n}\text{ a.s. for all } n\geq 1.
    \end{equation*}
\end{definition}
\begin{theorem}[Extended Ville's inequality; {\citealp[Theorem 4.1]{Wang2023}}]
    \label{thm:ExtVilleThm}
    Let $(X_{n})_{n\geq1}$ be an extended nonnegative supermartingale and let $c>0$. Then,
    \begin{equation*}
        \Pr\left(  \sup_{n\geq1}X_{n}\geq c\right)  \leq\mathbb{E}\left[\min\left(\frac{X_{1}}{c},1\right)\right].
    \end{equation*}
\end{theorem}

\subsection{(Extended) Ville mixture confidence sequences}
\label{sec:confidencesequenceglobal}

We now specialise the above inequalities to statistical inference.
Let $\Theta$ be the parameter space and, for each $\theta\in\Theta$, let $P_\theta$ be a probability law on the sequence $(Y_{i})_{i\geq1}$.
Assume that, for every $n\geq 1$, the law of $Y_{1:n}:=(Y_{1},\ldots,Y_{n})$ under $P_\theta$ is dominated by a fixed $\sigma$-finite measure $\lambda^{\otimes n}$, with density $\fn{n}{\theta}{y_{1:n}}>0$.
The data are generated under $P_{\thetatrue}$ for some fixed but unknown $\thetatrue\in\Theta$.

A confidence sequence $(C_{\alpha,n}(y_{1:n}))_{n\geq1}$ for $\thetatrue$ can be obtained by inverting a family of sequential tests of the point null hypotheses $H_{0,\theta_0}:\thetatrue=\theta_0$, $\theta_0\in\Theta$.
For each $\theta_0$, the method of mixtures yields an associated (extended) nonnegative martingale $(L_n(Y_{1:n};\theta_0))_{n\geq 1}$ under $H_{0,\theta_0}$, constructed as follows.
The approach dates back to the thesis of \citet{Ville1939} and was developed for confidence sequences by \citet{Darling1967}, \citet{Robbins1970b}, \citet{Robbins1970}, and \citet{Lai1976}.
\begin{assumption}
    \label{assump:finitemarginal}
    Let $\Pi_0$ be a $\sigma$-finite measure on $\Theta$ such that, for any $n\geq1$ and any $y_{1:n}\in\mathbb{R}^{n}$,
    \begin{equation*}
        \fintegral{n}{y_{1:n}}:=\int_{\Theta}\fn{n}{\theta}{y_{1:n}}\Pi_0(d\theta)<\infty.
    \end{equation*}
\end{assumption}
We refer to $\fintegral{n}{y_{1:n}}$ as the \emph{marginal likelihood} under $\Pi_0$, and to $\Pi_0$ itself as the \emph{prior}, since it encodes problem-specific information about $\theta$ into the procedure.
The prior is \emph{proper} if $\int_{\Theta}\Pi_0(d\theta)=1$ and \emph{improper} if $\int_{\Theta}\Pi_0(d\theta)=\infty$;
without loss of generality, any integrable $\Pi_0$ is taken to be proper.
\begin{remark}
    \label{rem:localprior}
    The confidence sequence derived by \citet{Robbins1970} is obtained by using a \textit{local} measure $\Pi_0(d\theta;\theta_0)$ that depends on, and is centred at, the tested value $\theta_0$;
    see \citet[Section 5.3 and Appendix D]{Wang2023}.
    While this choice leads to tractable closed-form expressions, the resulting CS is non-informative in the sense that it does not encode any prior belief about a user-specified global location, such as $0$.
    Except in \cref{sec:Gaussianimproper}, we restrict attention to priors that do not depend on $\theta_0$.
\end{remark}

For any $\theta_{0}\in\Theta$ and any $y_{1:n}\in\mathbb{R}^{n}$, define the mixture likelihood ratio
\begin{equation*}
    L_{n}(y_{1:n},\theta_{0})=\int_{\Theta}\frac{\fn{n}{\theta}{y_{1:n}}}{\fn{n}{\theta_{0}}{y_{1:n}}}\Pi_0(d\theta)=\frac{\fintegral{n}{y_{1:n}}}{\fn{n}{\theta_{0}}{y_{1:n}}}.
\end{equation*}
When $\Pi_0$ is proper, $L_{n}(y_{1:n},\theta_{0})$ is the reciprocal of the Bayes factor in favour of the point null $H_{0,\theta_0}:\thetatrue=\theta_0$ against the mixture alternative $H_1:\thetatrue=\theta,~~\theta\sim \Pi_0$ \citep{Shafer2011}. Under $H_{0,\theta_0}$, the process $(L_{n}(Y_{1:n},\theta_{0}))_{n\geq1}$ is then a nonnegative martingale with respect to the standard filtration, and $\mathbb{E}[L_{1}(Y_{1},\theta_{0})]=1$.
Hence, both \cref{thm:VilleCorollary,thm:ExtVilleThm} apply; see \citet{Lai1976} and \citet[Lemma 5.4]{Wang2023}.
When $\Pi_0$ is improper, $(L_{n}(Y_{1:n},\theta_{0}))_{n\geq1}$ is an extended nonnegative martingale under $H_{0,\theta_0}$, and $\mathbb{E}[L_{1}(Y_{1},\theta_{0})]$ need not be finite.
In that case, \cref{thm:VilleCorollary} no longer applies, while \cref{thm:ExtVilleThm} still does. Inverting these two inequalities yields two confidence-sequence constructions.

\begin{definition}[Ville (mixture) confidence sequence]
    Let $\Pi_0$ be a proper prior. The \emph{Ville (mixture) confidence sequence} (VCS) is the sequence of confidence regions $(\CV)_{n\geq1}$ defined by
    \begin{equation}
        \CVy=\left\{  \theta_{0}\mid L_{n}(y_{1:n},\theta_{0})\leq\frac{1}{\alpha}\right\}. \label{eq:VCS_generic}
    \end{equation}
\end{definition}

\begin{definition}[Extended Ville (mixture) confidence sequence]
\label{def:eVCS}
    Assume that, for each $\theta_0\in\Theta$, the random variable $L_1(Y_1,\theta_0)=m_1(Y_1)/f_{1,\theta_0}(Y_1)$ is continuous.
    For any $\theta_0\in\Theta$, let $g_{\theta_{0}}:(0,\infty)\to(0,1]$ be the monotone decreasing calibration function defined, for $c>0$ by
    \begin{equation}
        g_{\theta_0}(c)=\mathbb{E}_{Y_1\sim P_{\theta_0}}\left[  \min\left(  \frac{L_{1}(Y_{1},\theta_{0})}{c},1\right)  \right]\label{eq:gtheta}.
    \end{equation}
    Let $c_{\theta_0}^{\ast}=\inf(\supp L_1(Y_1,\theta_0))$. The restriction of $g_{\theta_0}$ to $(c_{\theta_0}^{\ast},\infty)$ is one-to-one, with inverse $g_{\theta_0}^{-1}:(0,1)\rightarrow(c_{\theta_0}^{\ast},\infty)$.
    The \emph{extended Ville (mixture) confidence sequence} (eVCS) is the sequence of confidence regions $(\CeV)_{n\geq 1}$ defined by
    \begin{equation}
        \CeVy=\left\{  \theta_{0}\mid L_{n}(y_{1:n},\theta_{0})\leq g_{\theta_{0}}^{-1}(\alpha)\right\}. \label{eq:eVCS_generic}
    \end{equation}
\end{definition}

    The names are justified by \cref{thm:VilleCorollary,thm:ExtVilleThm}, which imply that both sequences are valid $(1-\alpha)$-CSs for $\thetatrue$: for any $\thetatrue\in\Theta$ and any $\alpha\in(0,1)$,
    \begin{equation*}
        \Pr\left( \thetatrue\in \CVY \text{ for all }n\geq1\right) \geq 1-\alpha,
        \qquad
        \Pr\left(\thetatrue\in \CeVY \text{ for all }n\geq1\right) \geq 1-\alpha.
    \end{equation*}
    Since we only consider confidence sequences constructed by mixing likelihood ratios, we suppress the word ``mixture'' in the sequel.

\begin{remark}
When $\Pi_0$ is proper and admits a density, the mixture likelihood ratio can be interpreted as a prior-posterior ratio. Thus, the VCS and eVCS studied here can be viewed as level sets of a prior–posterior ratio martingale. Closely related constructions appear in the prior–posterior-ratio CSs of \citet{WaudbySmith2020}, in Gaussian-process uncertainty quantification \citep{Neiswanger2021}, in posterior-odds CSs for multinomial data \citep{Lindon2022}, and in the recent sequential-likelihood-mixing framework of \citet{Kirschner2025}.
\end{remark}

Note that, for any proper prior $\Pi_0$, $\alpha\in(0,1)$ and $\theta_0\in\Theta$, we have $g_{\theta_0}^{-1}(\alpha)\leq 1/\alpha$.
It follows that, for a given $\Pi_0$ and $\alpha$, the eVCS is necessarily tighter than the corresponding VCS. That is, for any $n\geq 1$, $y_{1:n}\in\bbR^n$,
\begin{equation*}
    \CeVy \subseteq \CVy.
\end{equation*}

We conclude this subsection by observing that every VCS confidence region is also a Bayesian credible region\footnote{Although this fact has not appeared in print, it has been known within the community since 2023 (personal communication with A.~Ramdas, H.~Wang, and P.~Gr\"unwald).}.
\begin{proposition}
    Let $\alpha\in(0,1)$, let $\Pi_0$ be a proper prior, and let $(\CVy)_{n\geq 1}$ be the corresponding VCS. Then, for any $n\geq 1$, $C_{\alpha,n}^{\mathrm{V}}(y_{1:n})$ is also a Bayesian $(1-\alpha)$ credible region:
    \begin{equation*}
        \int_{\theta\in \CVy } \frac{f_{n,\theta}(y_{1:n})\Pi_0(d\theta)}{m_n(y_{1:n})} \geq 1-\alpha.
    \end{equation*}
\end{proposition}
\begin{proof}
    $\int_{\theta\notin \CVy} \frac{f_{n,\theta}(y_{1:n})\Pi_0(d\theta)}{m_n(y_{1:n})} <  \alpha \int_{\theta\notin \CVy}\Pi_0(d\theta)\leq \alpha.$
\end{proof}

\subsection{Convexity of extended Ville confidence sequences}
\label{sec:convexity}

VCSs are convex regions (i.e., intervals when $\Theta \subseteq \bbR$) whenever $\Theta$ is convex and the likelihood $\fn{n}{\theta_0}{y_{1:n}}$ is log-concave in $\theta_0$.
This covers, for example, the scalar Gaussian mean model with known variance.
For eVCSs, convexity is less immediate because the threshold $g_{\theta_0}^{-1}(\alpha)$ in \cref{def:eVCS} also depends on $\theta_0$.
The following result provides sufficient conditions on the likelihood and marginal likelihood under which eVCSs are convex.

\begin{theorem}[Convexity of eVCSs]
\label{th:convexityeVCS}
Let $\Theta\subseteq \mathbb R^p$ be convex. For each $n\ge 1$, assume there exists a statistic $S_n(Y_{1:n})\in \mathcal S_n\subseteq \mathbb R^{q_n}$ such that
\[
L_n(y_{1:n};\theta)
=
\frac{\widetilde m_n(S_n(y_{1:n}))}{\widetilde f_{n,\theta}(S_n(y_{1:n}))},
\qquad
\widetilde m_n(s)
:=
\int_\Theta \widetilde f_{n,\vartheta}(s)\,\Pi_0(d\vartheta),
\]
where $\widetilde f_{n,\theta}$ is the density of $S_n$ under $P_\theta$.

Assume:
\begin{enumerate}
\item[(A1)] $(\theta,s)\mapsto \widetilde f_{1,\theta}(s)$ is jointly log-concave on $\Theta\times \mathcal S_1$;
\item[(A2)] for every $n\ge 1$ and every fixed $s\in\mathcal S_n$, the map $\theta\mapsto \widetilde f_{n,\theta}(s)$ is log-concave on $\Theta$;
\item[(A3)] $s\mapsto \widetilde m_1(s)$ is log-concave on $\mathcal S_1$.
\end{enumerate}

Then, for every $n\ge 1$, every $\alpha\in(0,1)$, and every dataset $y_{1:n}$, the eVCS $C_{\alpha,n}^{\mathrm{eV}}(y_{1:n})$ is a (possibly empty) convex subset of $\Theta$. When $p=1$, it is an interval.
\end{theorem}

Assumption (A3) is a condition on the joint behaviour of the likelihood and the prior.
The following proposition provides a simple sufficient condition on the prior for this to hold.

\begin{proposition}[Sufficient condition on the prior]
    If\/ $\Pi_0(d\theta)=\pi_0(\theta)\,d\theta$ with $\pi_0$ log-concave on $\Theta$, then in \cref{th:convexityeVCS}, Assumption \textnormal{(A3)} follows from Assumption \textnormal{(A1)}.
    \label{th:suffconditionprior}
\end{proposition}

Examples of probability laws $P_\theta$ for which Assumptions \textnormal{(A1)} and \textnormal{(A2)} are satisfied include
\begin{itemize}
\item[(i)] Gaussian mean models with known covariance;
\item[(ii)] location families of the form $f_{1,\theta}(y)=k(y-\theta)$ with log-concave kernel $k$ (e.g., Laplace, logistic, Gumbel);
\item[(iii)] Gaussian linear regression with fixed design and known covariance, taking $S_n=Y_{1:n}$.
\end{itemize}
In all these cases, eVCSs based on a log-concave prior are convex sets, and intervals in the scalar case.
The remainder of the paper focuses on case (i) in dimension one.

\section{Gaussian mean model and explicit confidence sequences}
\label{sec:Gaussian}

From this point onward, we specialise to the case of \iid Gaussian random variables $Y_i\sim\mathcal{N}(\thetatrue, \sigma^2)$ with unknown mean $\thetatrue$ and known variance $\sigma^2$, that is $\Theta=\bbR$ and
\[
  f_{n,\theta}(y_{1:n})=\prod_{i=1}^n \frac{1}{\sigma}\phi\left(\frac{y_i-\theta}{\sigma}\right),
\]
where $\phi$ denotes the standard normal density.
Since the sample mean $\bar{Y}_n = \tfrac{1}{n}\sum_{i=1}^n Y_i$ is a sufficient statistic for $\theta$, it is convenient to express both VCSs and eVCSs in terms of the likelihood and marginal likelihood densities of $\bar Y_n$:
\begin{equation}
    \fntilde{n,\theta}{\yn}   =\frac{1}{\sigma /\sqrt{n}} \phi\left(\frac{\overline y_n-\theta}{\sigma/\sqrt{n}}\right)\text{~~and~~~}\ftildeintegral{n}{\yn}=\int_\bbR\fntilde{n,\theta}{\yn}\Pi_0(d\theta).
    \label{eq:ftildenormal}
\end{equation}
Using \eqref{eq:ftildenormal}, the generic expressions \eqref{eq:VCS_generic} and \eqref{eq:eVCS_generic} specialise to the VCS
{\small
\begin{align}
\label{eq:Ville}
    \CVy &=\left\{  \theta_{0}\mid\frac{\ftildeintegral{n}{\yn}}{\fntilde{n,\theta_0}{\yn} }\leq\frac{1}{\alpha}\right\} =\left\{\theta_{0}=\yn-\delta\mid\delta^{2}\leq \frac{\sigma^{2}}{n}\log\left[  n\left(  \frac{1}{\alpha\sqrt{2\pi\sigma^{2}}\ftildeintegral{n}{\yn}}\right)  ^{2}\right]  \right\}\nonumber\\
    &=\left[\yn \pm \frac{\sigma}{\sqrt{n}} \sqrt{\log\left[  n\left(  \frac{1}{\alpha\sqrt{2\pi\sigma^{2}}\ftildeintegral{n}{\yn}}\right)  ^{2}\right]}     \right],
\end{align}}
and the eVCS
{\small
\begin{align}
\label{eq:extVille}
    \CeVy  &  =\left\{  \theta_{0}\mid\frac{\ftildeintegral{n}{\yn}}{\fntilde{n,\theta_0}{\yn} }\leq g_{\theta_0}^{-1}(\alpha)\right\}=\left\{  \theta_{0}=\yn-\delta\mid\delta^{2}\leq\frac{\sigma^{2}}{n}\log\left[  n\left(  \frac{g_{\yn-\delta}^{-1}(\alpha)}{\sqrt{2\pi\sigma^{2}}\ftildeintegral{n}{\yn}}\right)  ^{2}\right]\right\}.
\end{align}}

As previously mentioned, for scalar parameters the VCS \eqref{eq:Ville} is an interval, but the eVCS \eqref{eq:extVille} need not be.
For instance, \cref{fig:cs_disconnected} in \cref{sec:disconnected_cs} shows an example of a disconnected eVCS for the Gaussian mean model under a bimodal prior.
The following corollary of \cref{th:convexityeVCS,th:suffconditionprior} gives a simple sufficient condition on the prior under which the eVCS is an interval.

\begin{corollary}[Interval property of eVCSs in the Gaussian mean model]
\label{th:convexityeVCSGaussian}
If $\Pi_0$ has a log-concave density $\pi_0$ on $\mathbb R$, then the eVCS \eqref{eq:extVille} for the Gaussian mean is a (possibly empty) interval.
\end{corollary}

This covers, for example, Gaussian, Laplace, logistic, Gumbel, and normal-gamma priors (with shape parameter greater than $1$).
By contrast, horseshoe and Student-$t$ priors, as well as all other priors with power-law tails, fall outside this sufficient condition.
They are not excluded from the general framework, however, and will play a central role in \cref{sec:robustCS}.

The remainder of this section instantiates \eqref{eq:Ville} and \eqref{eq:extVille} for two representative choices of prior.
\Cref{sec:GaussianGaussianprior} treats the Gaussian prior, which yields a closed-form VCS and a semi-explicit eVCS representation, and is used in later sections as a running example of a prior that is \emph{not robust} to prior--data conflict.
\Cref{sec:Gaussianimproper} then introduces a one-parameter family of improper priors under which the calibration function $g_{\theta_0}$ admits a closed-form expression that is independent of $\theta_0$, and the eVCS reduces to an interval whose width does not depend on the data.
In \cref{sec:robustCS}, this second family arises as the limiting object of our main robustness theorem (\cref{thm:main}).

\subsection{VCS and eVCS under a Gaussian prior}
\label{sec:GaussianGaussianprior}

Consider the Gaussian prior $\Pi_0(d\theta)=\tfrac{1}{\tau}\phi\left(\frac{\theta-\mu}{\tau}\right)d\theta$ with mean $\mu\in\bbR$ and variance $\tau^2$.
This yields the VCS (\citealp{Pace2020,Pawel2024}, and \citealp[Section C.1]{Wang2023})
\begin{equation}
    \CVy=\left[  \yn\pm\frac{\sigma}{\sqrt{n}}\sqrt{\log\left(  1+\frac{\tau^{2}}{\sigma^{2}/n}\right) +\frac{(\yn-\mu)^{2}}{\sigma^{2}/n+\tau^{2}}-2\log\alpha}\right]
    \label{eq:VCS_gaussian}
\end{equation}
and the eVCS
\begin{equation}
    \CeVy=\left\{  \theta_{0}=\yn-\delta\mid\delta^2\leq  \frac{\sigma^2}{n}\left[\log\left(  1+\frac{\tau^{2}}{\sigma^{2}/n}\right) +\frac{(\yn-\mu)^{2}}{\sigma^{2}/n+\tau^{2}}+2\log g_{\yn-\delta}^{-1}(\alpha)\right]\right\}.
    \label{eq:eVCS_gaussian}
\end{equation}
Both are intervals by \cref{th:convexityeVCSGaussian}, since the Gaussian prior is log-concave.
The calibration function $g_{\theta_0}$ admits a closed-form expression, but its inverse $g_{\theta_0}^{-1}$ must be computed numerically.
In \cref{sec:robustCS}, we show that the volume of both confidence sequences diverges as the prior--data conflict $|\yn-\mu|$ grows.
This is the pathology that motivates our main robustness result (\cref{thm:main}).

\subsection{eVCS under an improper prior}
\label{sec:Gaussianimproper}

In contrast with Ville's inequality, the extended Ville's inequality accommodates improper priors.
For $\kappa\in\mathbb R$, consider the \emph{local} improper prior
\begin{equation}
\label{localimprop}
\Pi_0(d\theta;\theta_0)=\frac{1}{\sqrt{2\pi\sigma^{2}}}\exp\left(-\kappa(\theta-\theta
_{0})/\sigma\right)d\theta.
\end{equation}
We call this prior \emph{local} in the sense of \cref{rem:localprior}: when $\kappa\neq 0$, $\Pi_0$ depends on the tested value $\theta_0$;
when $\kappa=0$, $\Pi_0$ reduces to a constant and is independent of $\theta_0$.
The resulting likelihood ratio and calibration function are given by
\begin{align}
    L_{n}(y_{1:n};\theta_{0})  &=\frac{1}{\sqrt{n}}\exp\left(\frac{n}{2\sigma^{2}}\left(\overline{y}_{n}-\frac{\kappa\sigma}{n}-\theta_{0}\right)^{2}\right), \label{eq:Lnlocalimproper} \\
    g_{\theta_{0}}(c)& =\int_{-\infty}^{\infty}\min\left(  \frac{1}{c\sqrt{2\pi}\phi(u-\kappa)},1\right) \phi(u)du:=\widetilde{g}_{\kappa}(c) \label{eq:gkappadef}.
\end{align}
The key feature of this family is that the calibration function $g_{\theta_0}$ \eqref{eq:gkappadef} is \emph{independent} of $\theta_0$, and depends on the prior only through the scalar tail parameter $\kappa$.
We therefore write $\widetilde{g}_{\kappa}:[1,\infty)\to(0,1]$, and note that $\widetilde{g}_{\kappa}=\widetilde{g}_{-\kappa}$.
The function $\widetilde{g}_{\kappa}$ admits the following closed-form expression.

\begin{proposition}
    \label{thm:gkappa}
    We have
    \begin{equation}
        \label{eq:gtildekappa}
        \widetilde{g}_{\kappa}(x)=\left\{
        \begin{array}[c]{ll}
              1-\left[  \Phi\left(  \kappa + s(x)\right)  -\Phi\left(  \kappa-s(x)\right) \right] +\frac{\phi(\kappa-s(x))-\phi(\kappa+s(x))}{\kappa} & \kappa\neq0 \\
            2\left[  1-\Phi\left(  s(x) \right)  \right] + 2s(x)\phi(s(x))  & \kappa=0
        \end{array}
        \right.
    \end{equation}
    where $s(x)=\sqrt{\log x^2}$ and $\Phi$ is the cdf of the standard normal distribution. Moreover, $\widetilde{g}_{\kappa}$ is continuous, strictly decreasing, and one-to-one, with a continuous inverse $\widetilde{g}_{\kappa}^{-1}:(0,1]\to[1,\infty)$.
\end{proposition}

Substituting \eqref{eq:Lnlocalimproper} and \eqref{eq:gkappadef} into \cref{def:eVCS}, the resulting eVCS is given by
\begin{equation}
    \CeVy =\left[ \yn-\frac{\sigma\kappa}{n}\pm\frac{\sigma}{\sqrt{n}}\sqrt{\log\left(
n\widetilde{g}_{\kappa}^{-1}(\alpha)^{2}\right)  }\right].
    \label{eq:CeVlocalimproper}
\end{equation}
Two features of \eqref{eq:CeVlocalimproper} are worth emphasising.
First, unlike the Gaussian-prior VCS \eqref{eq:VCS_gaussian} and eVCS \eqref{eq:eVCS_gaussian}, the width of \eqref{eq:CeVlocalimproper} does \emph{not} depend on the observed data $y_{1:n}$.
Second, the centring relative to the sample mean is shifted by $-\sigma\kappa/n$, which encodes the asymptotically vanishing prior information carried by the exponential tilt in \eqref{localimprop}.

The special case $\kappa=0$ corresponds to the constant improper Jeffreys prior
\begin{equation}
    \label{eq:improperprior}
    \Pi_0(d\theta)=(2\pi\sigma^{2})^{-1/2}d\theta,
\end{equation}
where the multiplicative constant $(2\pi\sigma^{2})^{-1/2}$ is immaterial and chosen only for mathematical convenience.
The associated eVCS is
\begin{equation}
    \label{eq:CeVimproper}
    \CeVy =\left[ \yn\pm\frac{\sigma}{\sqrt{n}}\sqrt{\log\left(n\widetilde{g}_{0}^{-1}(\alpha)^{2}\right)  }\right],
\end{equation}
which coincides with the one introduced by \citet[Section 5.4]{Wang2023}.
On the other hand, the broader family \eqref{eq:CeVlocalimproper} indexed by $\kappa$ does not appear to have been used in this context.

\begin{remark}
The eVCS \eqref{eq:CeVimproper} admits a classical interpretation. With
\[
    a:=s\!\left(\widetilde g_0^{-1}(\alpha)\right),
    \qquad\text{so that}\qquad
    \alpha=\widetilde g_0\!\left(e^{a^2/2}\right)=2\left\{1-\Phi(a)+a\phi(a)\right\},
\]
the CS \eqref{eq:CeVimproper} can be rewritten as
\[
    \CeVy=\left[\yn \pm \frac{\sigma}{\sqrt n}\sqrt{a^2+\log n}\right],
\]
or, equivalently,
\[
    \theta_0\in\CeVY\ \forall n\ge 1
    \quad\Longleftrightarrow\quad
    \left|\sum_{i=1}^n\frac{Y_i-\theta_0}{\sigma}\right|
    \le
    \sqrt{n(a^2+\log n)}
    \ \forall n\ge 1.
\]
Hence, the eVCS under the improper constant prior coincides with Robbins' delayed-start Gaussian boundary with $m=1$ \citep[Eq.~(20)]{Robbins1970}, and with the corresponding Wiener-process boundary in \citet[p.~1414]{Robbins1970b} when $\tau=1$;
see also the related discussion in \citet[Section 2.6]{WaudbySmith2024}.
\end{remark}


\section{Bayes-assisted eVCSs with informative, bounded-influence priors}
\label{sec:robustCS}

This section establishes the main contribution of the paper.
We show that the eVCS associated with any proper prior having polynomial or exponential tails converges, as the prior--data conflict grows, to that of a matching improper prior from the family introduced in \cref{sec:Gaussianimproper}.
As a result, its volume remains uniformly bounded in the data, in sharp contrast with the VCS, whose volume diverges under \emph{every} proper prior.
\Cref{sec:priordataconflict} formalises prior--data conflict and establishes the VCS divergence; \cref{sec:mainresult} states the main robustness theorem; \cref{sec:proofsketch} sketches the proof.

\subsection{Prior--data conflict and divergence of the VCS}
\label{sec:priordataconflict}
For a proper prior $\Pi_0$, the marginal density $\ftildeintegral{n}{\cdot}$ of $\overline Y_n$ is the prior predictive density of the sufficient statistic.
Following \citet{Evans2006}, we quantify prior--data conflict by the prior-predictive tail probability
\begin{equation*}
    \tau_n(\yn,\Pi_0)
    :=
    \int_{\{t\in\bbR:\,\ftildeintegral{n}{t}\le \ftildeintegral{n}{\yn}\}}
    \ftildeintegral{n}{t}\,dt.
\end{equation*}
Small values of $\tau_n(\yn,\Pi_0)$ indicate that the observed $\yn$ lies in a low prior-predictive density region, and therefore signal a prior--data conflict.

When $\Pi_0$ is symmetric and unimodal around some location $\mu$, the prior predictive density $\ftildeintegral{n}{\cdot}$ is also symmetric and unimodal around $\mu$, as it is obtained by convolving $\Pi_0$ with a Gaussian density.
Consequently, $\tau_n(\yn,\Pi_0)$ depends on the data only through $|\yn-\mu|$, and is a nonincreasing function of $|\yn-\mu|$.
In particular,
\[
    \tau_n(\yn,\Pi_0)\to 0
    \qquad\text{as}\qquad
    |\yn-\mu|\to\infty.
\]
Thus, for symmetric unimodal priors, the magnitude of the prior--data conflict is naturally indexed by $|\yn-\mu|$.
Although the Evans--Moshonov criterion is only defined for proper priors, this motivates using $|\yn-\mu|$ as the conflict scale more generally, including for the improper priors considered below.

Finally, the Gaussian location model and the confidence regions constructed in this paper are shift-equivariant: replacing $Y_i$ by $Y_i-\mu$  and reparameterising $\theta$ to $\theta-\mu$ simply shifts the resulting confidence region by $\mu$. Therefore, without loss of generality, we take $\mu=0$ throughout and analyse the regime $\yn\to\pm\infty$.

The following result shows that, under any proper prior, the VCS becomes arbitrarily large in the presence of conflict.
\begin{theorem}
    \label{rk:Villepart}
    Let $\Pi_0$ be a proper prior. The associated VCS satisfies, for any $n\geq 1$,
    \[
	    \vol\left( C_{\alpha,n}^{\text{V}}(y_{1:n})\right)\rightarrow\infty\text{ as }\left\vert \yn\right\vert \rightarrow\infty.
    \]
\end{theorem}
\begin{proof}
Under a proper prior, $\ftildeintegral{n}{\yn}\to 0$ as $|\yn|\to\infty$.
Then, the result follows directly from the definition of VCS in \cref{eq:Ville}.
\end{proof}

\Cref{rk:Villepart} is independent of any tail behaviour of $\Pi_0$: informative and diffuse proper priors alike produce VCSs whose volume diverges as the conflict grows.
This is an intrinsic limitation of Ville's inequality, and a compelling reason to work with its extended form.
We now show that the extended Ville's inequality, combined with an appropriate tail condition on the prior, does permit uniform boundedness.

%
%

\subsection{Main robustness theorem for eVCS}
\label{sec:mainresult}

In order to obtain robust CSs in the presence of prior--data conflict, we require that the prior density have polynomial or exponential tails.
\begin{assumption}
\label{assump:tailprior}
Let $\Pi_0(d\theta)$ be a $\sigma$-finite prior distribution on $\bbR$ such that $\ftildeintegral{n}{z}$, defined by \cref{eq:ftildenormal}, is finite for any $z\in\bbR$.
Assume $\Pi_0$ has a density $\pi_0(\theta)$ with respect to the Lebesgue measure, finite almost everywhere, such that
\begin{equation}
    \pi_0(\theta)
    \sim
    \frac{C_{1}}{\sqrt{2\pi\sigma^{2}}}\left\vert \theta/\sigma\right\vert ^{-\beta}e^{-\kappa|\theta|/\sigma}
    \text{ as }
    \left\vert\theta\right\vert \rightarrow\infty
    \label{eq:powerlawpi}
\end{equation}
for some $C_1>0$, $\beta\geq0$ and $\kappa\geq0$.
\end{assumption}

When $\beta\leq1$ and $\kappa=0$, $\Pi_0$ is an improper prior; otherwise, the prior is proper.
Priors satisfying \cref{assump:tailprior} or similar tail conditions are known in the Bayesian robustness literature as \emph{bounded-influence priors} \citep{Dawid1973,Pericchi1992,Pericchi1995}, and include:
\begin{itemize}
    \item \emph{Improper constant} ($\kappa=\beta=0$): the constant improper prior \eqref{eq:improperprior} of \cref{sec:Gaussianimproper};
    \item \emph{Polynomial tails} ($\kappa=0$, $\beta>1$): proper priors such as the horseshoe \citep{Carvalho2010}, Student-$t$, and Cauchy priors;
    \item \emph{Exponential tails} ($\kappa>0$, $\beta=0$): proper priors such as the Laplace (double-exponential) prior \citep{Pericchi1992} and the normal-gamma prior \citep{Griffin2010}.
\end{itemize}

We can now state the main result.
Recall from \cref{sec:Gaussianimproper} that $\widetilde g_\kappa$ denotes the calibration function \eqref{eq:gkappadef} of the local improper prior \eqref{localimprop}, and that \eqref{eq:CeVlocalimproper} is the corresponding eVCS.

\begin{theorem}
\label{thm:main}
Let $\alpha\in(0,1)$, $\sigma>0$, and let $\Pi_0$ be a prior satisfying \cref{assump:tailprior} for some $\kappa\geq 0$.
Let $(\CeV)_{n\geq1}$ be the corresponding eVCS procedure.
For any fixed $n\geq 1$,
\begin{align*}
\CeVy- \overline
{y}_{n} &  \rightarrow\left [-\frac{\sigma\kappa}{n} \pm \frac{\sigma}{\sqrt{n}} \sqrt
{\log\left(  n\widetilde{g}_{\kappa}^{-1}(\alpha)^{2}\right)  }\right]\text{ as }\overline{y}%
_{n}\rightarrow \infty, \\
\CeVy- \overline
{y}_{n} &  \rightarrow\left [ \frac{\sigma\kappa}{n} \pm \frac{\sigma}{\sqrt{n}} \sqrt
{\log\left(  n\widetilde{g}_{\kappa}^{-1}(\alpha)^{2}\right)  }\right]  \text{ as }\overline{y}%
_{n}\rightarrow-\infty
\end{align*}
where the convergence is with respect to the Hausdorff distance on subsets of $\mathbb{R}$.
\end{theorem}

In words, under any proper prior satisfying the tail condition \eqref{eq:powerlawpi} with parameter $\kappa$, the eVCS converges to the eVCS \eqref{eq:CeVlocalimproper} corresponding to the $\kappa$-indexed member of the improper-prior family derived in \cref{sec:Gaussianimproper} as the prior--data conflict grows.
The width of the limiting interval \eqref{eq:CeVlocalimproper} is independent of the data, implying that, for fixed $n$, the volume of $\CeVy$ remains uniformly bounded over all $y_{1:n}\in\bbR^n$.
Furthermore, in the polynomial-tail case ($\kappa=0$), the limiting interval reduces to the constant-improper-prior eVCS \eqref{eq:CeVimproper} of \citet[Section 5.4]{Wang2023}.
That is, polynomial-tailed priors such as the horseshoe yield eVCSs that, under severe conflict, revert to the non-informative baseline. Exponential-tailed priors such as the Laplace retain bounded influence under severe prior--data conflict: the limiting interval is shifted toward the prior centre, but its width remains independent of the data.

\Cref{thm:main} is a statement about deterministic sequences $y_{1:n}$ with $\yn\to\pm\infty$.
The corresponding probabilistic statement, in which the conflict is induced by drawing data from a distant data-generating parameter, is given by the following corollary.
\begin{corollary}
\label{thm:corollary}
Under the assumptions of \cref{thm:main}, fix $n\geq1$ and, for each $\theta_0\in\bbR$, let $(Y_i^{(\theta_0)})_{i=1}^n$ be \iid $\mathcal N(\theta_0,\sigma^2)$.
Then, we have the following convergence in probability:
\begin{align*}
\CeVYtheta{\theta_0}- \overline{Y}_{n}^{(\theta_0) }&  \overset{Pr}{\rightarrow}\left [-\frac{\sigma\kappa}{n} \pm \frac{\sigma}{\sqrt{n}} \sqrt
{\log\left(  n\widetilde{g}_{\kappa}^{-1}(\alpha)^{2}\right)  }\right]\text{ as }\theta_0\rightarrow \infty, \\
\CeVYtheta{\theta_0}- \overline{Y}_{n}^{(\theta_0)} &  \overset{Pr}{\rightarrow}\left [ \frac{\sigma\kappa}{n} \pm \frac{\sigma}{\sqrt{n}} \sqrt
{\log\left(  n\widetilde{g}_{\kappa}^{-1}(\alpha)^{2}\right)  }\right]  \text{ as }\theta_0\rightarrow-\infty,
\end{align*}
where the convergence is with respect to the Hausdorff distance on subsets of $\mathbb{R}$.
\end{corollary}

Following a similar line of reasoning, it can be shown that, under any proper prior, the volume of $C_{\alpha,n}^{\mathrm V}(Y_{1:n}^{(\theta_0)})$ diverges in probability as $\vert\theta_0\vert\rightarrow\infty$.

\subsection{Sketch of the proof of  \cref{thm:main}}
\label{sec:proofsketch}

We give here a sketch of the proof when $\yn\rightarrow\infty$. Recall from \cref{eq:extVille} that the eVCS is of the form
\begin{align*}
\CeVy &  =\left\{  \theta_0=\yn-\delta\mid\delta^{2}\leq \frac{\sigma^{2}}%
{n}\log\left[  n\left(  \frac{g_{\yn-\delta}^{-1}(\alpha)}%
{\sqrt{2\pi\sigma^{2}}\ftildeintegral{n}{\yn}}\right)  ^{2}\right]
\right\},
\end{align*}
where $g_{\theta_0}^{-1}$ is the inverse of $g_{\theta_0}$, defined in \cref{eq:gtheta}. We aim to prove that, as $\yn\rightarrow\infty,$ one recovers the interval
$$\left [ \overline
{y}_{n}-\frac{\sigma\kappa}{n} \pm \frac{\sigma}{\sqrt{n}} \sqrt
{\log\left(  n\widetilde{g}_{\kappa}^{-1}(\alpha)^{2}\right)  }\right]$$
where $\widetilde{g}_{\kappa}^{-1}$ is the (continuous) inverse of the
continuous, strictly decreasing function $\widetilde{g}_{\kappa}%
:[1,\infty)\rightarrow(0,1]$ defined in \cref{eq:gtildekappa}. A key element of the proof is the following convergence result, proved in \cref{proofconv}.
\begin{proposition}
\label{thm:limitgtilde} For $\alpha\in(0,1)$, $\sigma>0$, and $\kappa\geq0$,
\begin{equation}
\lim_{\left\vert \theta\right\vert \rightarrow\infty}\frac{g_{\theta}%
^{-1}(\alpha)}{\pi_0(\theta)\sqrt{2\pi\sigma^{2}}}=\widetilde{g}_{\kappa}%
^{-1}(\alpha)
\end{equation}
\end{proposition}

The above result allows us to relate the asymptotic behaviour (in $\theta$) of $g_{\theta
}^{-1}(\alpha)$ to that of the prior $\pi_0(\theta)$. Additionally (\cref{prop:asympmarginal}), $\ftildeintegral{n}{z}$ has the same asymptotic behaviour, up to a constant, as $\pi_0(z)$. This allows us to relate the behaviour (in $\yn$) of $g_{\yn-\delta}^{-1}(\alpha)$ to that of $\ftildeintegral{n}{\yn}$ for any fixed $\delta$. The proof finally requires a uniform control over $\delta$.

\section{Point estimators for Ville and extended Ville CSs}
\label{sec:estimator}

In this section, we discuss the choice of the estimator to report alongside the VCS \eqref{eq:Ville} and eVCS \eqref{eq:extVille}. In the Gaussian mean model, the Ville region $\CVy$ always contains the sample mean $\yn$, whereas the extended Ville region $\CeVy$ need not. Moreover, even when $\yn\in\CeVy$, reporting $\yn$ ignores the prior information encoded by $\Pi_0$. We therefore propose to report instead the posterior mean
\begin{equation}
    \widehat\theta_n(\overline Y_n;\Pi_0)=\int_\bbR \theta \cdot\frac{\fntildebis{n,\theta}{\overline Y_n} \Pi_0(d\theta) }{\ftildeintegral{n}{\overline Y_n}}=\overline Y_n +\frac{\sigma^2}{n}\frac{{\widetilde{m}_n}'(\overline Y_n)}{\ftildeintegral{n}{\overline Y_n}},
\label{eq:BAestimator}
\end{equation}
which is the Bayesian estimator of the mean under the squared loss and the prior $\Pi_0$. Its expression in terms of the marginal likelihood of the sufficient statistic follows directly from Tweedie's formula; see, e.g., \citep{Efron2011}.
The following result states that, for any $n$, the estimate $\widehat\theta_n(\yn;\Pi_0)$ belongs to the corresponding extended Ville region, and thus also to the Ville region when $\Pi_0$ is proper.
\begin{proposition}
\label{prop:BAestimatorinCS}
For $n\geq 1$, $\alpha\in(0,1)$, $y_{1:n}\in\bbR^n$ and a (proper or improper) prior $\Pi_0$ satisfying Assumption \ref{assump:finitemarginal}, let  $\CeVy$ denote the eVCS in \cref{eq:extVille} for the Gaussian mean model.
The posterior mean $\widehat\theta_n(\overline y_n;\Pi_0)$ defined in \cref{eq:BAestimator} satisfies
\begin{equation}
    \widehat\theta_n(\yn;\Pi_0)\in \CeVy.
\end{equation}
If in addition, $\Pi_0$ is proper, then the VCS $\CVy$ defined by \eqref{eq:Ville} exists and $\CeVy \subseteq \CVy$, hence $\widehat\theta_n(\yn;\Pi_0)\in  \CVy$.
\end{proposition}

The proof of \cref{prop:BAestimatorinCS} builds on earlier work by \citet{Cortinovis2024} on properties of Pratt (fixed sample-size) confidence regions (CRs) \citep{Pratt1961,Pratt1963} for a Gaussian mean.
In particular, it exploits the fact that, for each $n$, the Pratt confidence region is contained in the corresponding extended Ville/Ville region. We also provide an alternative, direct proof for Ville CSs in \cref{app:ProofEstimatorVille}.
\begin{proof}
It is sufficient to prove that $\widehat\theta_n(\yn;\Pi_0)\in \CeVy$.
In our setting, the Pratt $(1-\alpha)$ confidence region \citep{Pratt1961,Pratt1963} is defined as
\begin{align}
\CPy :=\left \{ \theta_0 \mid \frac{\ftildeintegral{n}{\yn}}{\fntilde{n,\theta_0}{\yn}} \leq k_{n,\theta_0}(\alpha) \right \},
\label{eq:CPratt}
\end{align}
where $k_{n,\theta_0}(\alpha)$ is the $(1-\alpha)$ quantile of $T_{n,\theta_0}(\Yn)=\frac{\ftildeintegral{n}{\Yn}}{\fntilde{n,\theta_0}{\Yn}}$ under $\thetatrue=\theta_0$. Recall that the eVCS is of the form
\begin{align}
\CeVy:=\left \{ \theta_0 \mid \frac{\ftildeintegral{n}{\yn}}{\fntilde{n,\theta_0}{\yn}} \leq g_{\theta_{0}}^{-1}(\alpha) \right \}
\end{align}
and satisfies $\Pr(\thetatrue\in \CeVY)\geq 1-\alpha.$
Therefore, by definition of $k_{n,\theta_0}(\alpha)$ in \cref{eq:CPratt}, one necessarily has $k_{n,\theta_0}(\alpha)\leq g_{\theta_{0}}^{-1}(\alpha)$. It follows that for any $\alpha\in(0,1)$, $n\geq 1$, any $y_{1:n}\in\bbR^n$ and any prior $\Pi_0$,
$$
\CPy \subseteq \CeVy.
$$
\citet{Cortinovis2024} have shown that $\widehat\theta_n(\yn;\Pi_0)\in \CPy$ for any $n\geq 1$, $\alpha\in(0,1)$, and any non-degenerate prior $\Pi_0$, with the degenerate case being covered by \citet{Pratt1961}. \cref{prop:BAestimatorinCS} then follows.
\end{proof}

The implication of \cref{prop:BAestimatorinCS} can be visualised by means of $p$-value functions~\citep{Fraser1991,Schweder2016}, which, for a given confidence procedure $C_{\alpha,n}(y_{1:n})$, are defined as
\begin{equation*}
    p_{y_{1:n}}(\theta_0) = \sup\{\alpha \in (0, 1) \mid \theta_0 \in C_{\alpha,n}(y_{1:n})\},
\end{equation*}
allowing one to visualise nested confidence regions across the whole confidence range $\alpha \in (0, 1)$. In our case the Ville $p$-value $p^{\mathrm{V}}_{y_{1:n}}(\theta_0)$ and extended Ville $p$-value $p^{\mathrm{eV}}_{y_{1:n}}(\theta_0)$ are
\begin{align}
p^{\mathrm{V}}_{y_{1:n}}(\theta_0) & =\min\left(1,1/ L_{n}(y_{1:n},\theta_0)\right),\\
p^{\mathrm{eV}}_{y_{1:n}}(\theta_0) & =g_{\theta_0}(L_{n}(y_{1:n},\theta_0)).
\end{align}

\begin{figure}
    \centering
    \includegraphics[width=\textwidth]{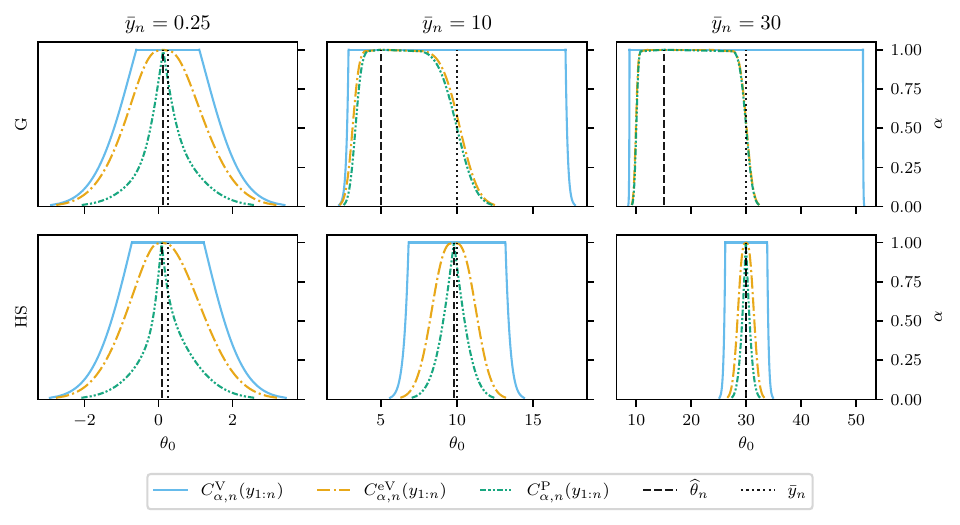}
    \caption{$p$-value functions for the Ville, extended Ville and Pratt CR procedures for the mean of a Gaussian with variance $\sigma^2=1$ under a Gaussian (G, first row) and horseshoe (HS, second row) prior with location $0$ and scale $1$, for $n = 1$ and when observing $\bar y_n \in \{0.25, 10, 30\}$.}
        \label{fig:estimator_cc}
\end{figure}

\cref{fig:estimator_cc} compares the $p$-value functions of the Ville, extended Ville and Pratt confidence regions under zero-centred Gaussian and horseshoe priors.
As expected,
$$
p^{\mathrm{P}}_{y_{1:n}}\leq p^{\mathrm{eV}}_{y_{1:n}}\leq p^{\mathrm{V}}_{y_{1:n}}.
$$
Furthermore, the posterior mean $\widehat\theta_n(\bar y_n; \Pi_0)$ falls within all the confidence regions for every $\alpha \in (0, 1)$, justifying its use as a point estimator for confidence sequences constructed by the method of mixtures. Note that the $p$-value functions associated with our confidence regions are themselves capped e-posteriors \citep[Figure 1 and Definition 4.1]{Gruenwald2023}.

\begin{figure}
    \centering
    \includegraphics[width=\textwidth]{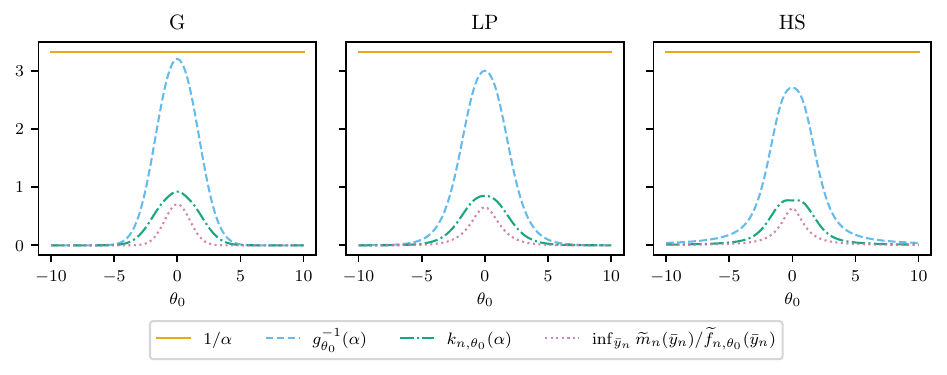}
    \caption{Sublevel set thresholds used to construct the Ville ($1/\alpha$), extended Ville ($g_{\theta_0}^{-1}(\alpha)$) and Pratt ($k_{n, \theta_0}(\alpha)$) CRs for the mean of a Gaussian with variance $\sigma^2=1$ under Gaussian (G), Laplace (LP) and horseshoe (HS) priors with location $0$ and scale $1$, for $n = 1$ and $\alpha = 0.3$.}
        \label{fig:estimator}
\end{figure}

As discussed in its proof, \cref{prop:BAestimatorinCS} follows from the specific hierarchy existing among the sublevel set thresholds used to construct the Ville, extended Ville and Pratt confidence regions.
These quantities are illustrated in \cref{fig:estimator} for three proper priors.
In all cases, we have $1 / \alpha \geq g_{\theta_0}^{-1}(\alpha) \geq k_{n,\theta_0}(\alpha)$, with the additional trivial lower bound $c_{\theta_0}^*=\inf_{\bar y_n} \ftildeintegral{n}{\yn} / \widetilde f_{n,\theta_0}(\bar y_n)$.

\section{Simulation studies}
\label{sec:exp}

We present numerical studies illustrating the theoretical properties established in \cref{sec:robustCS}.
As elsewhere in the article, we consider \iid Gaussian observations with known variance $\sigma^2 = 1$.
We compare confidence sequences obtained under the following zero-centred priors, each with scale parameter one: Gaussian ($\mathrm{G}$), Laplace ($\mathrm{LP}$), horseshoe ($\mathrm{HS}$), and Student-$t$ with 5 degrees of freedom ($\mathrm{T}_5$).
For the extended Ville CS, we also include the improper constant prior ($\mathrm{I}$) discussed in \cref{sec:Gaussianimproper}.

\begin{figure}[ht]
    \centering
    \includegraphics[width=\textwidth]{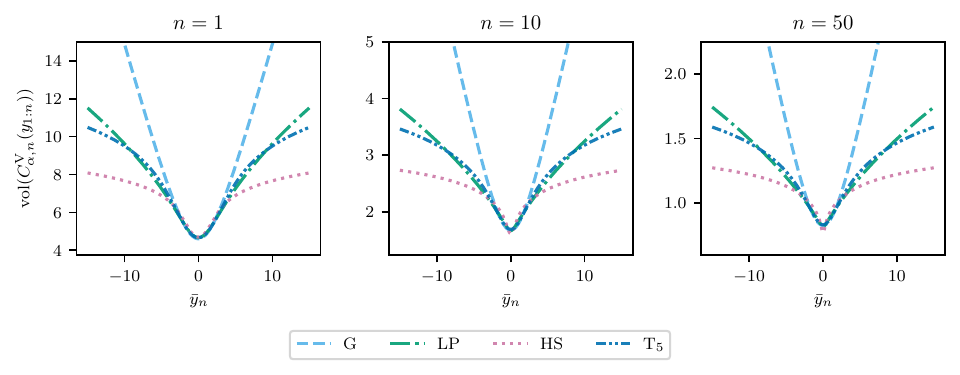}
    \caption{VCS volume under different priors as a function of $\bar y_n$ for $n \in \{1, 10, 50\}$ and $\alpha = 0.1$.  G=Gaussian, LP=Laplace, HS=Horseshoe, $\mathrm{T}_5$=Student-$t$ with 5 degrees of freedom.}
    \label{fig:ville_volume}
\end{figure}

\cref{fig:ville_volume} shows the volume of the Ville confidence region at fixed sample size $n$ as a function of $\bar y_n$, for three choices of $n$.
For the VCS, all proper priors exhibit the same qualitative behaviour: the confidence region is shortest when the prior--data conflict is small ($\bar y_n \simeq 0$), but its volume diverges as $|\bar y_n|$ grows, in accordance with \cref{rk:Villepart}.
Notably, heavier-tailed priors exhibit a slower divergence rate in this experiment.

\begin{figure}[ht]
    \centering
    \includegraphics[width=\textwidth]{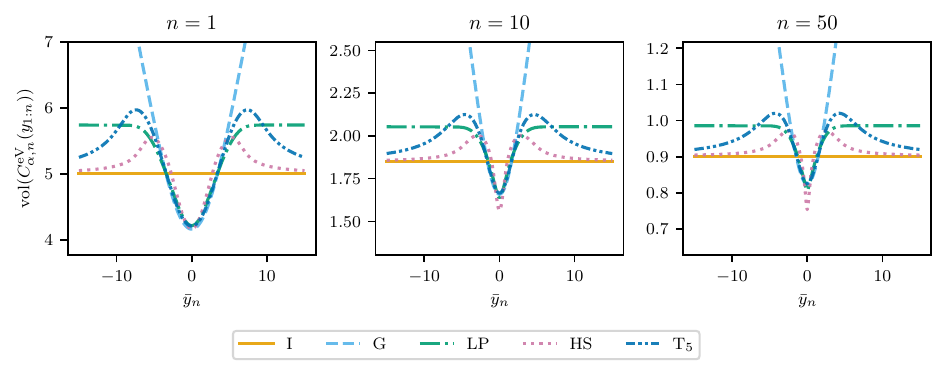}
    \caption{eVCS volume under different priors as a function of $\bar y_n$ for $n \in \{1, 10, 50\}$ and $\alpha = 0.1$. I=Improper, G=Gaussian, LP=Laplace, HS=Horseshoe, $\mathrm{T}_5$=Student-$t$ with 5 degrees of freedom.}
    \label{fig:eville_volume}
\end{figure}

By contrast, \cref{fig:eville_volume} shows that the extended Ville CS exhibits a markedly different pattern across priors in the presence of prior--data conflict, as described by \cref{thm:main}.
The Gaussian prior still leads to confidence regions whose volume diverges as $|\bar y_n|$ grows.
On the other hand, the Laplace, horseshoe, and Student-$t$ priors satisfy \cref{assump:tailprior} and thus yield, for each fixed $n$, confidence regions whose volume remains bounded as a function of $\bar y_n$.
Moreover, for the polynomial-tailed priors considered here, namely the horseshoe and Student-$t$ priors, the extended Ville CSs revert to the non-informative improper-prior CS as the prior--data conflict grows. 
Overall, heavy-tailed priors satisfying \cref{assump:tailprior} exploit prior information while exhibiting robustness to prior--data conflict.

\section{Discussion}
\label{sec:discussion}

Under \cref{assump:tailprior}, \cref{thm:postasymp} implies that, for each fixed $n$, the $(1-\alpha)$ highest-posterior-density (HPD) credible interval $C^{\text{Bayes}}_{\alpha,n}(y_{1:n})$ satisfies
\begin{align}
C^{\text{Bayes}}_{\alpha,n}(y_{1:n})-\yn \to \left[-\frac{\sigma\kappa}{n}\pm \frac{\sigma}{\sqrt{n}}z_{1-\alpha/2} \right ]
\label{eq:convergencecredible}
\end{align}
in Hausdorff distance as $\yn\to\infty$.
The corresponding limit as $\yn\to-\infty$ is obtained by replacing $-\sigma\kappa/n$ with $\sigma\kappa/n$;
see also \citet{Dawid1973,Pericchi1992,Pericchi1995}.
Under this lens, the result in \cref{thm:main} can be seen as the anytime-valid analogue of \cref{eq:convergencecredible}: under the same tail condition, the eVCS converges to an interval with the same asymptotic centre but a larger half-width, and thus eventually fully contains the limiting HPD credible interval.
Furthermore, \citet{Cortinovis2024} derived a result parallel to \cref{thm:main} for Pratt confidence regions (CRs) \citep{Pratt1961,Pratt1963}, which are briefly discussed in \cref{sec:estimator}.
Together, these results place anytime-valid confidence sequences, HPD credible intervals, and Pratt's Bayes-optimal confidence regions within the same asymptotic robustness framework.

Convexity of eVCSs (\cref{sec:convexity}) remains only partially understood.
In our numerical studies, eVCSs were found to be intervals for all symmetric unimodal priors considered, including the horseshoe and Student-$t$ priors.
At the same time, the example in \cref{sec:disconnected_cs} shows that disconnected eVCSs can occur under asymmetric mixture priors.
Characterising the class of priors for which eVCSs are intervals, or generalising the sufficient condition in \cref{th:suffconditionprior} beyond log-concavity, is therefore an important open problem.

While this article focuses on Gaussian observations with known variance, the underlying ideas are not intrinsically tied to this setting.
\citet{WaudbySmith2024} recently introduced asymptotic confidence sequences as anytime-valid counterparts of CLT-based confidence intervals in broad nonparametric settings.
Their construction also relies on combining the method of mixtures with Ville-type inequalities, suggesting that bounded-influence priors could play a similar robustness-enhancing role beyond the exact Gaussian model.
Extending \cref{thm:main} to such asymptotic nonparametric settings is therefore a particularly promising direction for future work.

Lastly, a number of other extensions beyond the scalar Gaussian mean model, including multivariate models, general location families, and sub-Gaussian observation models, would broaden the method's applicability and warrant further investigation.

\newpage

\appendix


\section{Proofs}
\label{sec:omittedproof}

\subsection{Proofs of \cref{sec:ville}}
\label{sec:convexityproof}

\paragraph{Proof of \cref{th:convexityeVCS}.}
Fix $n\ge 1$ and $\alpha\in(0,1)$. Let $y_{1:n}$ be some data with statistic $s_n\in\mathcal S_n$. Since $g_{\theta_0}$ is decreasing, we have
\[
\theta_0\in \CeVy
\quad\Longleftrightarrow\quad
g_{\theta_0}(L_n(y_{1:n},\theta_0))\ge \alpha.
\]
Define $$p_n(\theta_0):=g_{\theta_0}(L_n(s_n,\theta_0)).$$ Then
$\CeVy=\{\theta_0 \mid p_n(\theta_0)\ge \alpha\}$. Thus it suffices to show that $p_n(\theta_0)$ is log-concave in $\theta_0$ as superlevel sets of a log-concave function are convex.
For $c>0$, the calibration function is
\begin{align*}
g_{\theta_0}(c)
&=\mathbb E_{S\sim \fntilde{1,\theta_0}{\cdot}}\!\left[\min\!\left(\frac{\ftildeintegral{1}{S}}{c\,\fntilde{1,\theta_0}{S}},\,1\right)\right]
=\int_{\mathcal S_1} \min\!\left(\frac{\widetilde m_1(s)}{c},\ \widetilde f_{1,\theta_0}(s)\right)\,ds.
\end{align*}
Plugging $c=L_n(y_{1:n},\theta_0)$ yields
\[
p_n(\theta_0)
=\int_{\mathcal S_1} \min\!\left(\widetilde f_{1,\theta_0}(s),\ a_n(\theta_0)\,\widetilde m_1(s)\right)\,ds.
\]
where
\[
a_n(\theta_0):=\frac{1}{L_n(y_{1:n},\theta_0)}=\frac{\fntilde{n,\theta_0}{s_n}}{\ftildeintegral{n}{s_n}}.
\]
Let
\[
u(\theta_0,s):=\widetilde f_{1,\theta_0}(s),\qquad v(\theta_0,s):=a_n(\theta_0)\,\widetilde m_1(s),\qquad h(\theta_0,s):=\min\{u(\theta_0,s),v(\theta_0,s)\}.
\]
Then
$$
p_n(\theta_0)=\int_{\mathcal S_1} h(\theta_0,s)\,ds.$$

By Assumption (A1), $u(\theta_0,s)$ is jointly log-concave. By Assumption (A2), $a_n(\theta_0)$ is log-concave in $\theta_0$, and by Assumption (A3), $\widetilde m_1(s)$ is log-concave in $s$. Hence
$$
\log v(\theta_0,s)=\log a_n(\theta_0) + \log\widetilde m_1(s)
$$
is jointly concave, so $v$ is jointly log-concave. Since
$$
\log h = \min(\log u, \log v)
$$
and the pointwise minimum of concave functions is concave, $h$ is jointly log-concave. By Pr\'ekopa's theorem, marginalisation preserves log-concavity, so $p_n(\theta_0)=\int_{\mathcal S_1} h(\theta_0,s)\,ds$ is log-concave. Therefore, its superlevel sets are convex.

\paragraph{Proof of \cref{th:suffconditionprior}.}

If $\Pi_0(d\theta)=\pi_0(\theta)d\theta$ with $\pi_0$ log-concave then Assumption (A1) implies that
$$
(\theta,s)\mapsto \widetilde f_{1,\theta}(s)\pi_0(\theta)
$$
is jointly log-concave. Hence, by Pr\'ekopa's theorem,
$$\widetilde m_1(s)=\int_\Theta \widetilde f_{1,\theta}(s)\pi_0(\theta)d\theta$$
is log-concave.

\subsection{Proofs of \cref{sec:Gaussian}}
\paragraph{Proof of \cref{thm:gkappa}.}
    We have
   {\small
        \begin{align*}
            \widetilde{g}_{\kappa}(x)  &  =\frac{1}{x\sqrt{2\pi}}\int_{\phi(u-\kappa)>1/(x\sqrt{2\pi})}\frac{\phi(u)}{\phi(u-\kappa)}du+\int_{\phi(u-\kappa)<1/(x\sqrt{2\pi})}\phi(u)du \\
            &  =\frac{1}{x\sqrt{2\pi}}e^{\frac{\kappa^{2}}{2}}\int_{\left\vert u-\kappa\right\vert <\sqrt{\log(x^{2})}}e^{-u\kappa}du+1-\int_{\left\vert u-\kappa\right\vert <\sqrt{\log(x^{2})}}\phi(u)du \\
            &= \left\{
            \begin{array}[c]{ll}
                \frac{e^{-\kappa^2/2}}{x\sqrt{2\pi}\kappa}\left(  e^{\kappa\sqrt{\log(x^{2})}}-e^{-\kappa\sqrt{\log(x^{2})}}\right)  +1-\left[  \Phi\left(  \kappa +\sqrt{\log(x^{2})}\right)  -\Phi\left(  \kappa-\sqrt{\log(x^{2})}\right)\right]  & \kappa\neq0\\
                \frac{2}{x\sqrt{2\pi}}\sqrt{\log(x^{2})}+2\left[  1-\Phi\left(  \sqrt{\log(x^{2})}\right)  \right]  & \kappa=0.
            \end{array}
            \right.
    \end{align*}}
From this analytical expression, it is obvious that $\widetilde{g}_{\kappa}$ is continuous. To establish the strict decrease, it is enough to notice that if $1\leq x_1 \leq x_2$, then
\[
\min\left( \frac{1}{x_2\sqrt{2\pi}\phi(u-\kappa)},1\right) \leq \min\left( \frac{1}{x_1\sqrt{2\pi}\phi(u-\kappa)},1\right)
\]
with the inequality being strict when $\phi(u-\kappa)>1/(x\sqrt{2\pi})$, which is of positive measure. Integrating both sides with respect to $\phi(u)du$ shows the function is strictly decreasing, hence one-to-one.

\subsection{Proofs of \cref{sec:robustCS}}
\subsubsection{Proof of \cref{thm:main}}

We prove the result for $\yn\rightarrow\infty$. The case
$\yn\rightarrow-\infty$ proceeds similarly.

\paragraph{Preliminary results:}

We first state asymptotic properties of the marginal likelihood and
posterior density under a prior satisfying \cref{assump:tailprior}. Similar results are derived by~\citet{Dawid1973,Pericchi1992} and  \citet{Pericchi1995} under related assumptions.

\begin{proposition}
\label{prop:asympmarginal}
Under Assumption \ref{assump:tailprior}, the marginal likelihood of the sufficient statistic satisfies
\[
\ftildeintegral{n}{\yn}\sim\frac{C_{1}e^{\frac{\kappa^{2}}{2n}}}%
{\sqrt{2\pi\sigma^{2}}}\left\vert \yn/\sigma\right\vert ^{-\beta
}e^{-\frac{\kappa\left\vert \yn\right\vert }{\sigma}}\text{ as
}\left\vert \yn\right\vert \rightarrow\infty\text{.}%
\]
\end{proposition}
Let $\pi_n(\theta|\yn)$ be the posterior probability density function of $\theta$ given $\overline{Y}_n=\yn$ evaluated at $\theta$, defined by
$$
\pi_n(\theta|\yn)=\frac{\fntilde{n,\theta}{\yn}\pi_0(\theta)}{\ftildeintegral{n}{\yn}}.
$$

\begin{proposition}
\label{thm:postasymp}
Under Assumption \ref{assump:tailprior}, the posterior density satisfies, for any $\theta\in\mathbb{R},$
\begin{align*}
\pi_n(\yn-\frac{\kappa\sigma}{n}-\theta &  \mid \yn%
)\rightarrow\frac{\sqrt{n}}{\sqrt{2\pi\sigma^{2}}}e^{-\frac{n\theta^{2}%
}{2\sigma^{2}}}\text{ as }\yn\rightarrow\infty,\\
\pi_n(\yn+\frac{\kappa\sigma}{n}-\theta &  \mid \yn%
)\rightarrow\frac{\sqrt{n}}{\sqrt{2\pi\sigma^{2}}}e^{-\frac{n\theta^{2}%
}{2\sigma^{2}}}\text{ as }\yn\rightarrow-\infty.
\end{align*}
In particular, for any $z\in\mathbb{R}$,
\[
\pi_n(\theta\mid\theta+z)\rightarrow\left\{
\begin{array}
[c]{ll}%
\fntilde{n,\kappa\sigma/n}{z} & \text{as }\theta\rightarrow\infty\\
\fntilde{n,-\kappa\sigma/n}{z} & \text{as }\theta\rightarrow-\infty.
\end{array}
\right.
\]

\end{proposition}

\paragraph{Plan of the proof:}

Recall from \cref{eq:extVille} that the eVCS is of the form
\begin{align*}
\CeVy &  =\left\{  \theta_0=\yn-\delta\mid\delta^{2}\leq \frac{\sigma^{2}}%
{n}\log\left[  n\left(  \frac{g_{\yn-\delta}^{-1}(\alpha)}%
{\sqrt{2\pi\sigma^{2}}\ftildeintegral{n}{\yn}}\right)  ^{2}\right],
\right\}
\end{align*}
where $g_{\theta_0}^{-1}$ is the inverse of $g_{\theta_0}$, defined in \cref{eq:gtheta}. We aim to prove that, as $\yn\rightarrow\infty,$ one recovers the interval
$$\left [ \overline
{y}_{n}-\frac{\sigma\kappa}{n} \pm \frac{\sigma}{\sqrt{n}} \sqrt
{\log\left(  n\widetilde{g}_{\kappa}^{-1}(\alpha)^{2}\right)  }\right]$$
where $\widetilde{g}_{\kappa}^{-1}$ is the (continuous) inverse of the
continuous, strictly decreasing function $\widetilde{g}_{\kappa}%
:[1,\infty)\rightarrow(0,1]$ defined in \cref{eq:gtildekappa}. A key element of the proof is the following convergence result, proved in \cref{proofconv}.
\begin{proposition}
\label{thm:limitgtilde} For $\alpha\in(0,1)$, $\sigma>0$, and $\kappa\geq0$,
\begin{equation}
\lim_{\left\vert \theta\right\vert \rightarrow\infty}\frac{g_{\theta}%
^{-1}(\alpha)}{\pi_0(\theta)\sqrt{2\pi\sigma^{2}}}=\widetilde{g}_{\kappa}%
^{-1}(\alpha)
\end{equation}
\end{proposition}

The above result allows us to relate the asymptotic behaviour (in $\theta$) of $g_{\theta
}^{-1}(\alpha)$ to that of the prior $\pi_0(\theta)$ and, therefore, in light of \cref{prop:asympmarginal}, the behaviour (in $\yn$) of $g_{\yn-\delta}^{-1}(\alpha)$ to that of $\ftildeintegral{n}{\yn}$ for any fixed $\delta$. The proof finally requires a uniform control over $\delta$ and is split into three main steps.

\begin{enumerate}
\item First, we prove that, for any $\xi\in(0,1)$, there exists $T_{1}>1$ such that, for all $\overline{y}_{n}>T_{1}$, we have
$$
    C^{\text{eV}}_{\alpha,n}(y_{1:n})\subseteq\left\lbrack\yn-\xi\yn,\yn+\xi\overline{y}_{n}\right\rbrack.
$$
\item Then, we prove that there exists $\delta_{\max}>0$ such that, for all
$\yn>T_{1}$, $$C^{\text{eV}}_{\alpha,n}(y_{1:n})\subseteq\left\lbrack\overline
{y}_{n}-\delta_{\max},\yn+\delta_{\max}\right\rbrack.$$
\item Finally, we prove that, for any $\varepsilon>0$, there exists $T_{2}>0$ such
that, for all $\yn>T_{2}$,
\end{enumerate}
{\small
\begin{align*}
    &\left[\pm \left(
   \frac{\sigma}{\sqrt{n}} \sqrt{\log\left[  n\left(  \widetilde{g}_{\kappa}^{-1}(\alpha)\right)^{2}\right]}    -\varepsilon
    \right)\right]
    \subseteq C^{\text{eV}}_{\alpha,n}(\yn)-(\yn-\frac{\kappa\sigma}{n}) \subseteq
    \left[\pm \left(
   \frac{\sigma}{\sqrt{n}} \sqrt{\log\left[  n\left(  \widetilde{g}_{\kappa}^{-1}(\alpha)\right)^{2}\right]}    +\varepsilon
    \right)\right].
\end{align*}}

The proof relies on the fact that \cref{eq:powerlawpi} implies that the function $\pi_0(\theta)e^{\kappa\theta/\sigma}$ is a regularly varying function at infinity. Such functions roughly behave as power functions and satisfy a number of properties; see \citet{Bingham1987} and \cref{app:regularvariation} for background material.

\paragraph{Proof of step 1:}

\begin{proposition}
For any $\xi\in(0,1)$, there exists $T_{1}>1$ such that, for all $\overline
{y}_{n}>T_{1}$, we have $$\CeVy\subseteq
\left\lbrack\yn-\xi\yn,\yn+\xi\overline
{y}_{n}\right\rbrack.$$
\label{prop:mainproofstep1}
\end{proposition}

\begin{proof}
Let
$$
A_{\yn}(\delta)=\frac{\sigma^{2}}{n}\log\left[  n\left(  \frac{g_{\yn-\delta}^{-1}(\alpha)}{\sqrt{2\pi\sigma^{2}}\ftildeintegral{n}{\yn}}\right)  ^{2}\right]
$$
and $A^+_{\yn}(\delta)=\max(A_{\yn}(\delta),0)$. By \cref{eq:extVille}, if $\theta_0=\yn-\delta\in \CeVy$, then $\delta^2\leq A_{\yn}(\delta)\leq A^+_{\yn}(\delta)$. So it is enough to prove that, for any $\xi\in(0,1)$,
\begin{align}
\sup_{\delta||\delta|>\xi\yn}\frac{A^+_{\yn}(\delta)}{\delta^{2}} \longrightarrow0\text{ as
}\yn\rightarrow\infty.
\end{align}
We will split the supremum between the two cases $\yn\geq \delta$ and $\yn\leq\delta$. We have (\cref{thm:limitgtilde})
\[
\lim_{\theta\rightarrow\infty}\frac{g_{\theta}^{-1}(\alpha)}{\sqrt{2\pi
\sigma^{2}}\pi_0(\theta)}=\widetilde{g}_{\kappa}^{-1}(\alpha).
\]
Using \cref{eq:powerlawpi}, it follows that $g_{\theta}^{-1}(\alpha)e^{\frac{\kappa\theta}{\sigma}}$ is a
regularly varying function of $\theta$ at infinity. Hence, for any $\theta\in
[0,\infty)$, $g_{\theta}%
^{-1}(\alpha)$ admits the representation
\[
g_{\theta}^{-1}(\alpha)=\ell_{1}(\theta)(1+\theta)^{-\beta}e^{-\frac{\kappa\theta
}{\sigma}},
\]
where $\ell_{1}$ is a slowly varying function that converges to a positive constant.
Similarly, for any $z\in [0,\infty)$, $\ftildeintegral{n}{z}$ admits the representation
\[
\sqrt{2\pi\sigma^{2}}\ftildeintegral{n}{z}=\ell_{2}(z)(1+z)^{-\beta}e^{-\frac{\kappa
z}{\sigma}},
\]
where $\ell_{2}$ is a slowly varying function that converges to a positive constant. Then, for $\yn \geq \max(0,\delta)$,
{\small
\begin{align}
A_{\yn}(\delta)
& =\frac{\sigma^{2}}{n}\left(  \log n+\log\left(  \frac{\ell
_{1}(\yn-\delta)^{2}}{\ell_{2}(\yn)^{2}}\right)
+2\beta\log\left( \frac{1+\yn}{1+\yn-\delta}\right)  +\frac
{2\delta\kappa}{\sigma}\right) \label{eq:regvarbound}\\
&\leq \frac{\sigma^{2}}{n}\left(  \log n+\log\left(  \frac{\ell
_{1}(\yn-\delta)^{2}}{\ell_{2}(\yn)^{2}}\right)
+2\beta\log\left( 1+\yn\right)  +\frac
{2\delta\kappa}{\sigma}\right) \nonumber \\
&\leq \frac{\sigma^{2}}{n}\left(  \log n+M
+2\beta\log\left( 1+\yn\right)  +\frac
{2\delta\kappa}{\sigma}\right),\nonumber
\end{align}}
where $M$ is an upper bound for $\log\left(  \frac{\ell
_{1}(\yn-\delta)^{2}}{\ell_{2}(\yn)^{2}}\right)$, which exists as both $\ell_1$ and $\ell_2$ converge to a positive constant. It follows that
\[
\sup_{\delta\mid |\delta|>\xi \yn,\delta\leq \yn}\frac{A^+_{\yn}(\delta)}{\delta^{2}}  \longrightarrow0\text{ as
}\yn\rightarrow\infty.
\]

Similarly, for any $\theta\in(-\infty,0]$, we have
\[
g_{\theta}^{-1}(\alpha)=\tilde\ell_{1}(-\theta)(1-\theta)^{-\beta}e^{\frac{\kappa\theta
}{\sigma}},
\]
where $\tilde\ell_{1}$ is a slowly varying function that converges to a positive constant. Then, for $0\leq \yn\leq \delta$, we have

{\small
\begin{align*}
A_{\yn}(\delta) &=\frac{\sigma^{2}}{n}\left(  \log n+\log\left(  \frac{\tilde\ell
_{1}(-(\yn-\delta))^{2}}{\ell_{2}(\yn)^{2}}\right) +2\beta\log\left( \frac{1+\yn}{1-\yn+\delta}\right)  +\frac{2\delta\kappa}{\sigma}\right) \\
&\leq \frac{\sigma^{2}}{n}\left(  \log n+\log\left(  \frac{\tilde\ell
_{1}(-(\yn-\delta))^{2}}{\ell_{2}(\yn)^{2}}\right)
+2\beta\log\left( 1+\yn\right)  +\frac
{2\delta\kappa}{\sigma}\right)\\
&\leq \frac{\sigma^{2}}{n}\left(  \log n+\widetilde{M}
+2\beta\log\left( 1+\yn\right)  +\frac
{2\delta\kappa}{\sigma}\right),
\end{align*}}

where $\widetilde{M}$ is an upper bound for $\log\left(  \frac{\tilde\ell
_{1}(-(\yn-\delta))^{2}}{\ell_{2}(\yn)^{2}}\right)$, which exists as both $\tilde\ell_1$ and $\ell_2$ converge to a positive constant. Hence,
\[
\sup_{\delta\mid |\delta|>\xi\yn,\delta\geq\yn}\frac{A^+_{\yn}(\delta)}{\delta^{2}} \longrightarrow0\text{ as
}\yn\rightarrow\infty.
\]
Hence, there exists $T_1>0$ such that, for all $\yn>T_1,\vert\delta\vert>\xi\yn$, we have $\overline
{y}_{n}-\delta\notin \CeVy$.
\end{proof}

\paragraph{Proof of step 2:}

\begin{proposition}
There exists $\delta_{\max}>0$ such that, for all $\yn>T_{1}$,
$$C^{\text{eV}}_{\alpha,n}(y_{1:n})\subseteq\left\lbrack\yn-\delta_{\max}
,\yn+\delta_{\max}\right\rbrack.$$
\end{proposition}

\begin{proof}
By \cref{prop:mainproofstep1}, for any $\xi\in(0,1)$, there is $T_{1}>0$ such
that, when $\yn>T_{1}%
$, $\theta=\yn-\delta\in C^{\text{eV}}_{\alpha,n}(\yn)$ implies%
\[
-\xi\yn\leq\delta\leq\xi\yn%
\]
hence%
\[
\yn-\delta\geq\yn(1-\xi)>(1-\xi)T_{1}\geq0.
\]

Therefore, \cref{eq:regvarbound} holds. As already stated, $\ell_{1}$ and $\ell_{2}$ are bounded away from
0 and infinity (continuous on $[1,\infty)$ and converging to a constant). Hence,
\[
\log\left(  n\frac{\ell_{1}(\yn-\delta
)^2}{\ell_{2}(\yn)^2}\right)  \leq  M%
\]
for some $M>0$. Additionally, since $- \xi\yn\leq \delta\leq \xi\yn$, then $-\xi<\frac{\delta}{\yn +1}<\xi$ and
\begin{align}
-2\beta
\log(1+\xi)<-2\beta\log\left(\frac{1+\yn-\delta}{1+\yn}\right)  <-2\beta
\log(1-\xi).
\end{align}
By combining the two inequalities above with \cref{eq:regvarbound}, it follows that there exists $\delta_{\max}>0$ (which does not depend on $\yn$) such that, for all $|\delta|>\delta_{\max}$, $\yn>T_1$,
\[
\frac{A_{\yn}(\delta)}{\delta^2}\leq
\frac{\sigma^2}{n}\frac{M-2\beta\log (1-\xi) +\frac{2\delta\kappa}{\sigma}}{\delta^{2}}<1.
\]
Hence, for all $|\delta|>\delta_{\max}$, $\yn>T_{1}$, $\theta
=\yn-\delta\notin \CeVy$, from which the claim follows.
\end{proof}

\paragraph{Proof of step 3:}

To finish the proof, we can work, for $\yn>T_{1},$ on the compact
set $[-\delta_{\max},\delta_{\max}]$. Using the Uniform Convergence Theorem for regularly varying functions (see \citealp[Theorem 1.2.1]{Bingham1987} and \cref{thm:UCT}),
\[
\frac{g_{\yn-\delta}^{-1}(\alpha)}{g_{\yn}%
^{-1}(\alpha)}\rightarrow e^{\frac{\kappa\delta}{\sigma}}\text{ as }%
\yn\rightarrow\infty\text{, uniformly for }\delta\in
\lbrack-\delta_{\max},\delta_{\max}]\text{ }%
\]
Additionally, since $\ftildeintegral{n}{z}\sim\pi_0(z)e^{\frac{\kappa^{2}}{2n}}$ as
$z\rightarrow\infty$, we have
\[
\frac{g_{\yn}^{-1}(\alpha)}{\sqrt{2\pi\sigma^{2}}\ftildeintegral{n}{\yn}}\rightarrow \widetilde{g}_{\kappa}^{-1}(\alpha) e^{-\frac{\kappa^{2}}{2n}}\text{ as }%
\yn\rightarrow\infty\text{.}%
\]
It follows that
$$
\frac{g_{\yn-\delta}^{-1}(\alpha)}{\sqrt{2\pi\sigma^{2}%
}\ftildeintegral{n}{\yn}}=\frac{g_{\yn-\delta}%
^{-1}(\alpha)}{g_{\yn}^{-1}(\alpha)}\frac{g_{\yn%
}^{-1}(\alpha)}{\sqrt{2\pi\sigma^{2}}\ftildeintegral{n}{\yn}}\rightarrow\widetilde{g}_{\kappa}^{-1}(\alpha)e^{\frac{\kappa\delta}{\sigma
}-\frac{\kappa^{2}}{2n}}$$
as $\yn\rightarrow\infty$, uniformly for $\delta\in\lbrack-\delta_{\max},\delta_{\max}\rbrack$.
Hence, for any $\varepsilon>0$, there exists $T_{2}>0$ such that, for all
$\yn>T_{2}$, $|\delta|<\delta_{\max}$,
\begin{align*}
    \frac{2\kappa\sigma\delta}{n}-\frac{\sigma^{2}\kappa^{2}}{n^{2}}+\frac
    {\sigma^{2}}{n}\log\left[  n\left(  \widetilde{g}_{\kappa}^{-1}(\alpha
    )\right)  ^{2}\right]  -\varepsilon &\leq\frac{\sigma^{2}}{n}\log\left[
    n\left(  \frac{g_{\yn-\delta}^{-1}(\alpha)}{\sqrt{2\pi\sigma^{2}
    }\ftildeintegral{n}{\yn}}\right)  ^{2}\right]  \\
    &\leq\frac{2\kappa
    \sigma\delta}{n}-\frac{\sigma^{2}\kappa^{2}}{n^{2}}+\frac{\sigma^2}{n
    }\log\left[  n\left(  \widetilde{g}_{\kappa}^{-1}(\alpha)\right)  ^{2}\right]
    +\varepsilon .
\end{align*}
It follows that
\begin{align*}
    &\left\{  \theta=\yn-\delta\mid\left(  \delta-\frac{\kappa\sigma
    }{n}\right)  ^{2}\leq\frac{\sigma^{2}}{n}\log\left[  n\left(  \widetilde{g}%
    _{\kappa}^{-1}(\alpha)\right)  ^{2}\right]  -\varepsilon\right\} \\
    &\quad\ \subseteq
    \CeVy \subseteq\left\{  \theta=\yn-\delta
    \mid\left(  \delta-\frac{\kappa\sigma}{n}\right)  ^{2}\leq\frac{\sigma^{2}}%
    {n}\log\left[  n\left(  \widetilde{g}_{\kappa}^{-1}(\alpha)\right)
    ^{2}\right]  +\varepsilon\right\}
\end{align*}
and, therefore, $\CeVy-\yn$ converges, in Hausdorff distance as $\yn\to\infty$, to the interval
\[
\left[  -\frac{\kappa\sigma
}{n} \pm \frac{\sigma}{\sqrt{n}}\sqrt{\log\left[  n\left(  \widetilde{g}%
_{\kappa}^{-1}(\alpha)\right)  ^{2}\right]} \right].
\]

\subsubsection{Proofs of the preliminary results}

\paragraph{Proof of \cref{prop:asympmarginal}.}

We have
\begin{align*}
\ftildeintegral{n}{\yn}  &  =\frac{1}{\sqrt{2\pi\sigma^{2}/n}}\int
e^{-\frac{(\theta-\yn)^{2}}{2\sigma^{2}/n}}\pi_0(\theta)d\theta\\
&  =\frac{1}{\sqrt{2\pi\sigma^{2}/n}}\int e^{-\frac{n(\theta-\overline{y}%
_{n})^{2}}{2\sigma^{2}}-\frac{2n(\kappa\sigma\theta/n)}{2\sigma^{2}}}%
\pi_0(\theta)e^{\frac{\kappa\theta}{\sigma}}d\theta\\
&  =\frac{1}{\sqrt{2\pi\sigma^{2}/n}}e^{-\frac{\yn^{2}%
-(\yn-\kappa\sigma/n)^{2}}{2\sigma^{2}/n}}\int e^{-\frac
{n(\theta-\yn+\kappa\sigma/n)^{2}}{2\sigma^{2}}}\pi_0
(\theta)e^{\frac{\kappa\theta}{\sigma}}d\theta\\
&  =e^{\frac{\kappa^{2}}{2n}}e^{-\frac{\kappa\yn}{\sigma}}%
\frac{1}{\sqrt{2\pi\sigma^{2}/n}}\int e^{-\frac{n(\theta-\overline{y}%
_{n}+\kappa\sigma/n)^{2}}{2\sigma^{2}}}\pi_0(\theta)e^{\frac{\kappa\theta
}{\sigma}}d\theta
\end{align*}
Now, $\pi_0(\theta)e^{\frac{\kappa\theta}{\sigma}}\sim\frac{C_{1}}{\sqrt
{2\pi\sigma^{2}}}(\theta/\sigma)^{-\beta}$ as $\theta\rightarrow\infty$, and is bounded in $-\infty$. Using \cref{thm:RVGauss}, we have
\[
\frac{1}{\sqrt{2\pi\sigma^{2}/n}}\int e^{-\frac{n(\theta-\overline{y}%
_{n}+\kappa\sigma/n)^{2}}{2\sigma^{2}}}\pi_0(\theta)e^{\frac{\kappa\theta
}{\sigma}}d\theta\sim\frac{C_{1}}{\sqrt{2\pi\sigma^{2}}}(\overline{y}%
_{n}/\sigma)^{-\beta}\text{ as }\yn\rightarrow\infty.
\]
It follows that, as $\yn\rightarrow\infty$,
\[
\ftildeintegral{n}{\yn}\sim e^{\frac{\kappa^{2}}{2n}}e^{-\frac
{\kappa\yn}{\sigma}}\frac{C_{1}}{\sqrt{2\pi\sigma^{2}}}%
(\yn/\sigma)^{-\beta}.
\]
The proof for $\yn\rightarrow-\infty$ proceeds similarly.

\paragraph{Proof of \cref{thm:postasymp}.}

Recall that
\[
\ftildeintegral{n}{z}\sim\frac{C_{1}e^{\frac{\kappa^{2}}{2n}}}{\sqrt{2\pi\sigma
^{2}}}\left\vert z/\sigma\right\vert ^{-\beta}e^{-\frac{\kappa\left\vert
z\right\vert }{\sigma}}\text{ as }\left\vert z\right\vert \rightarrow\infty.
\]
We have, as $z\rightarrow\infty$%
\begin{align*}
\frac{\pi_0(z-\frac{\kappa\sigma}{n}-\theta)}{\ftildeintegral{n}{z}}  &  \sim
\frac{\frac{C_{1}}{\sqrt{2\pi\sigma^{2}}}\left\vert (z-\frac{\kappa\sigma}%
{n}-\theta)/\sigma\right\vert ^{-\beta}e^{-\kappa(z-\frac{\kappa\sigma}%
{n}-\theta)/\sigma}}{\frac{C_{1}e^{\frac{\kappa^{2}}{2n}}}{\sqrt{2\pi
\sigma^{2}}}\left\vert z/\sigma\right\vert ^{-\beta}e^{-\frac{\kappa z}%
{\sigma}}}  \rightarrow e^{\frac{\kappa^{2}}{2n}}e^{\frac{\kappa\theta}{\sigma}}.
\end{align*}
Hence,
\begin{align*}
\pi_n(\yn-\frac{\kappa\sigma}{n}-\theta\mid \yn)
&  =\frac{\pi_0(\yn-\frac{\kappa\sigma}{n}-\theta)\fntilde{n,\yn-\frac{\kappa\sigma}{n}-\theta}{\yn}}{\ftildeintegral{n}{\yn}}\\
&  =\frac{\pi_0(\yn-\frac{\kappa\sigma}{n}-\theta)\fntilde{n,0}{\theta+\frac{\kappa\sigma}{n}}}{\ftildeintegral{n}{\yn} }\\
&  \rightarrow e^{\frac{\kappa^{2}}{2n}}e^{\frac{\kappa\theta}{\sigma}%
}\fntilde{n,0}{\theta+\frac{\kappa\sigma}{n}}=\fntilde{n,0}{\theta}\text{ as }\yn\rightarrow\infty.
\end{align*}
Similarly,
\[
\pi_n\left(  \yn+\frac{\kappa\sigma}{n}-\theta\mid \yn\right)
\rightarrow\fntilde{n,0}{\theta}\text{ as }\yn\rightarrow
-\infty
\].

For the final claim, let \(x\to\infty\). Then
\[
\pi_n(x\mid x+z)
=
\pi_n\!\left((x+z)-\frac{\kappa\sigma}{n}-\Bigl(z-\frac{\kappa\sigma}{n}\Bigr)\,\middle|\,x+z\right)
\to
\fntilde{n,0}{z-\frac{\kappa\sigma}{n}}
=
\fntilde{n,\kappa\sigma/n}{z}.
\]
Similarly, as \(x\to-\infty\),
\[
\pi_n(x\mid x+z)\to \fntilde{n,-\kappa\sigma/n}{z}.
\]

\paragraph{Proof of \cref{thm:limitgtilde}.}
\label{proofconv}

We prove the result for $\theta\rightarrow\infty$. The case
$\theta\rightarrow-\infty$ proceeds similarly. The proof uses the following lemma.

\begin{lemma}
Let $c>0$ and $\Pi_0$ be some prior that satisfies Assumption \ref{assump:tailprior} for some $\kappa\geq 0$. Let $g_\theta$ be the associated calibration function. Then
$$
\lim_{\theta\rightarrow\infty}g_{\theta}\left(  c\sqrt{2\pi\sigma^{2}}%
\pi_0(\theta)\right)  =\widetilde{g}_{\kappa}(c).
$$
\label{lemma:gtheta}
\end{lemma}

\begin{proof}
Let $K=\frac{1}{c\sqrt{2\pi\sigma^{2}}}$. We have, for $Y_1\sim f_{1,\theta}$,
\begin{align*}
g_{\theta}\left(  \frac{\pi_0(\theta)}{K}\right)   &  =\mathbb{E}\left[
\min\left(  \frac{Km_1(Y_{1})}{\pi_0(\theta)\fn{1}{\theta}{Y_{1}}},1\right)  \right] =\mathbb{E}\left[  \min\left(  \frac{K}{\pi_1(\theta\mid Y_{1})},1\right)
\right] \\
&  =\mathbb{E}\left[  1_{\pi_1(\theta\mid Y_{1})>K}\frac{K}{\pi_1(\theta\mid
Y_{1})}\right]  +\mathbb{E}\left[  1_{\pi_1(\theta\mid Y_{1})<K}\right] \\
&  =1-\mathbb{E}\left[  1_{\pi_1(\theta\mid Y_{1})>K}\left(  1-\frac{K}%
{\pi_1(\theta\mid Y_{1})}\right)  \right].
\end{align*}
For any $z\in\mathbb{R},$ the posterior satisfies (see \cref{thm:postasymp})
\[
\pi_1(\theta\mid z+\theta)\rightarrow \fntilde{1,\kappa\sigma}{z}=\frac{1}{\sqrt
{2\pi\sigma^{2}}}e^{-\frac{(z-\kappa\sigma)^{2}}{2\sigma^{2}}}\text{ as
}\theta\rightarrow\infty\text{. }%
\]
By dominated convergence, noting that $1_{\pi_1(\theta\mid y)>K}\left(
1-\frac{K}{\pi_1(\theta\mid y)}\right)  \in\lbrack0,1]$,
\begin{align*}
\mathbb{E}\left[  1_{\pi_1(\theta\mid Y_{1})>K}\left(  1-\frac{K}{\pi(\theta\mid
Y_{1})}\right)  \right]   &  =\int1_{\pi_1(\theta\mid Y_{1})>K}\left(
1-\frac{K}{\pi_1(\theta\mid Y_{1})}\right)  f_{1,\theta}(y_{1})dy_{1}\\
&  =\int1_{\pi_1(\theta\mid z+\theta)>K}\left(  1-\frac{K}{\pi_1(\theta\mid
z+\theta)}\right)  f_{1,0}(z)dz\\
&  \rightarrow\int1_{f_{1,0}(z-\kappa\sigma)>K}\left(  1-\frac{K}{f_{1,0}%
(z-\kappa\sigma)}\right)  f_{1,0}(z)dz\text{ as }\theta\rightarrow\infty,
\end{align*}
where one can check that one minus the RHS limit is $\widetilde{g}_{\kappa
}(c)=\widetilde{g}_{\kappa}\left(  \frac{1}{K\sqrt{2\pi\sigma^{2}}}\right)  $.
\end{proof}

We now prove \cref{thm:limitgtilde}. Let $0<\varepsilon<\min(\alpha,1-\alpha)$. Let $c_{1}=\widetilde{g}_\kappa
^{-1}(\alpha-\varepsilon)>0$ and $c_{2}=\widetilde{g}_\kappa^{-1}
(\alpha+\varepsilon)>0$. From \cref{lemma:gtheta}, we have
\[
\lim_{\theta\rightarrow\infty}g_{\theta}\left(  c_{i}\sqrt{2\pi\sigma^{2}}%
\pi_0(\theta)\right)  =\widetilde{g}_{\kappa}(c_{i}), \quad i=1,2.
\]
Hence, there exists $A>0$ such that, for all $\theta>A$,
\[
\widetilde{g}_{\kappa}(c_{i})-\varepsilon\leq g_{\theta}\left(  c_{i}%
\sqrt{2\pi\sigma^{2}}\pi_0(\theta)\right)  \leq\widetilde{g}_{\kappa}%
(c_{i})+\varepsilon, \quad i=1,2
\]
and, as $g_{\theta}^{-1}$ is monotone decreasing (see \cref{thm:gkappa}),
\[
\frac{g_{\theta}^{-1}\left(  \widetilde{g}_\kappa(c_{i})+\varepsilon\right)  }%
{\sqrt{2\pi\sigma^{2}}\pi_0(\theta)}\leq c_{i}\leq\frac{g_{\theta}^{-1}\left(
\widetilde{g}_\kappa(c_{i})-\varepsilon\right)  }{\sqrt{2\pi\sigma^{2}}\pi_0(\theta)}\text{, }i=1,2.
\]
It follows that
\[
\widetilde{g}_{\kappa}^{-1}(\alpha+\varepsilon
)\leq\frac{g_{\theta}^{-1}\left(  \alpha\right)  }{\sqrt{2\pi\sigma^{2}}%
\pi_0(\theta)}\leq\widetilde{g}_{\kappa}^{-1}(\alpha-\varepsilon).
\]
Finally, as $\widetilde{g}_\kappa^{-1}$ is a continuous function (see \cref{thm:gkappa}), we conclude that
\[
\lim_{\theta\rightarrow\infty}\frac{g_{\theta}^{-1}\left(  \alpha\right)
}{\sqrt{2\pi\sigma^{2}}\pi_0(\theta)}=\widetilde{g}_{\kappa}^{-1}(\alpha).
\]

\subsubsection{Proof of \cref{thm:corollary}}

We prove the result for $\theta_0 \rightarrow \infty$; the case $\theta_0 \rightarrow -\infty$ proceeds similarly. Let $d_H$ denote the Hausdorff distance. To simplify the notation, we introduce the following:

\begin{align*}
C(y_{1:n}) &= \CeVy - \overline{y}_{n} \\
I &= \left[ -\frac{\sigma\kappa}{n} \pm \frac{\sigma}{\sqrt{n}} \sqrt{\log(n\widetilde{g}_{\kappa}^{-1}(\alpha)^{2})} \right]
\end{align*}

Let $\epsilon > 0$. By \cref{thm:main}, there exists $M_\epsilon > 0$ such that for all $y_{1:n}$, we have:
\[
\overline{y}_n > M_\epsilon \Rightarrow d_H(C(y_{1:n}), I) < \epsilon,
\]
which is equivalent to:
\[
d_H(C(y_{1:n}), I) \geq \epsilon \Rightarrow \overline{y}_n \leq M_\epsilon.
\]

This leads to the following inclusion of events:
\[
\left\{ d_H(C(Y^{(\theta_0)}_{1:n}), I) \geq \epsilon \right\} \subset \left\{ \overline{Y}^{(\theta_0)}_n \leq M_\epsilon \right\},
\]
which immediately implies:
\[
\mathbb{P}\left( d_H(C(Y^{(\theta_0)}_{1:n}), I) \geq \epsilon \right) \leq \mathbb{P}\left( \overline{Y}^{(\theta_0)}_n \leq M_\epsilon \right).
\]

As $\theta_0 \rightarrow \infty$, we have $\overline{Y}^{(\theta_0)}_n \xrightarrow{Pr} \infty$. This implies that $\mathbb{P}\left( \overline{Y}^{(\theta_0)}_n \leq M_\epsilon \right) \rightarrow 0$ and, consequently, $\mathbb{P}\left( d_H(C(Y^{(\theta_0)}_{1:n}), I) \geq \epsilon \right) \rightarrow 0$. Since this holds for any $\epsilon > 0$, we conclude that:
\[
d_H(C(Y^{(\theta_0)}_{1:n}), I) \xrightarrow{Pr} 0.
\]

\subsection{Proofs of \cref{sec:estimator}}
\paragraph{Direct proof of \cref{prop:BAestimatorinCS} for Ville CSs:}
\label{app:ProofEstimatorVille}

It is sufficient to prove that, for any $n\geq 1$, $z\in\bbR$,
\begin{align}
\left |\frac{\sigma^2}{n}\frac{\widetilde{m}'_{n}(z)}{\ftildeintegral{n}{z}}\right| \leq \sqrt{\frac{\sigma^{2}}{n}\log\left[  n\left(\frac{1}{\sqrt{2\pi\sigma^{2}}\ftildeintegral{n}{z}}\right)^{2}\right]}.
\label{eq:posterior_eq0}
\end{align}
Let $\theta\sim \Pi_0$ and assume $Z\mid \theta \sim \mathcal{N}(\theta,\sigma^2/n)$. Then the conditional distribution of $\theta$ given $Z=z$ is
$$
\Pi_n(d\theta \mid z)=\frac{\fntilde{n,\theta}{z}\Pi_0(d\theta)}{\ftildeintegral{n}{z}}.
$$
By Tweedie's formula~\citep{Efron2011}, we have
\begin{align}
\bbE[\theta- z\mid Z=z]=\frac{\sigma^2}{n}\frac{\widetilde{m}'_{n}(z)}{\ftildeintegral{n}{z}}.
\label{eq:posterior_eq1}
\end{align}
By the Cauchy-Schwarz inequality,
\begin{align}
\left |\bbE[\theta - z\mid Z=z]\right|\leq \sqrt{\bbE[(\theta - z)^2 \mid Z=z]}.
\label{eq:posterior_eq2}
\end{align}
The Kullback--Leibler divergence between the posterior $\Pi_n( \cdot \mid z)$ and the prior $\Pi_0(\cdot)$ is
\begin{align*}
0\leq \KL(\Pi_n(\cdot\mid z)~\|~\Pi_0(\cdot))&=\bbE[\log \fntilde{n,\theta}{z} \mid Z=z]- \log\ftildeintegral{n}{z}\\
&=\frac{1}{2}\log \frac{n}{2\pi\sigma^2\ftildeintegral{n}{z}^2}-\frac{n}{2\sigma^2}\bbE[(\theta - z)^2 \mid Z=z].
\end{align*}
Hence
\begin{align}
\bbE[(\theta - z)^2 \mid Z=z]\leq \frac{\sigma^2}{n}\log \frac{n}{2\pi\sigma^2\ftildeintegral{n}{z}^2}.
\label{eq:posterior_eq3}
\end{align}
Combining the inequalities \eqref{eq:posterior_eq3} and \eqref{eq:posterior_eq2} with \cref{eq:posterior_eq1},  we obtain the inequality \eqref{eq:posterior_eq0}.


\section{Background material on regularly varying functions}
\label{app:regularvariation}

This section provides additional background and secondary results on regularly varying functions \citep{Bingham1987}.

\subsection{Definitions}
\begin{definition}[Slowly varying function]
A function $\ell:[0,\infty)\to(0,\infty)$ is slowly varying at infinity if for all $c>0$,
$$
\frac{\ell(cx)}{\ell(x)}\to 1\text{  as }x\to\infty.
$$
\end{definition}
\noindent Examples of slowly varying functions include $\log^a$, for $a\in\Real$, and functions converging to a constant $c>0$, .

\begin{definition}[Regularly varying function]
A function $h:[0,\infty)\to(0,\infty)$ is regularly varying at infinity with $\rho\in\Real$ if $h(x)=x^\rho\ell(x)$ for some slowly varying function $\ell$. $\rho$ is called the index of variation.
\end{definition}

\subsection{Uniform Convergence Theorem}

\begin{proposition}
(\citealp{Bingham1987}, Theorem 1.2.1). Let $\ell$ be a slowly varying function
defined on $[c,\infty)$, for some $c>0$. Then, for any $0<a\leq b<\infty$,%
\[
\sup_{\lambda\in\lbrack a,b]}\left\vert \frac{\ell(\lambda x)}{\ell
(x)}-1\right\vert \rightarrow0\text{ as }x\rightarrow\infty.
\]

\end{proposition}

We have the following direct corollary.

\begin{proposition}
\label{thm:UCT}
Let $h(x)=\ell(x)x^{\rho}$ be a regularly varying function on $[c,\infty)$,
for some $c>0$, with index of variation $\rho\in\mathbb{R}$. Then, for any
$-\infty<a\leq b<\infty$,
\[
\sup_{y\in\lbrack a,b]}\left\vert \frac{h(x-y)}{h(x)}-1\right\vert
\rightarrow0\text{ as }x\rightarrow\infty.
\]

\end{proposition}

\begin{proof}
$\ell(x-y)=\ell(x(1-\frac{y}{x}))$ where, for $x>x_{0}=\max(|a|,|b|)+1$,
$0<1-\frac{b}{x_{0}}\leq1-\frac{y}{x}\leq2$. By the previous proposition,
$\frac{\ell(x-y)}{\ell(x)}\rightarrow1$ uniformly for $y\in\lbrack a,b]$. For
the power-law part, $\left(  1-\frac{y}{x}\right)  ^{\rho}$ is bounded between
$\left(  1-\frac{a}{x}\right)  ^{\rho}$ and $\left(  1-\frac{b}{x}\right)
^{\rho}$.\ So by sandwiching, it converges uniformly to 1 on $y\in\lbrack
a,b]$. Hence $\frac{h(x-y)}{h(x)}\rightarrow1$ as $x\rightarrow\infty$
uniformly for $y\in\lbrack a,b]$.
\end{proof}

\subsection{Potter's theorem and convolution with a Gaussian pdf}

\begin{proposition} (Potter's theorem, \citet[Theorem 1.5.6]{Bingham1987})
\label{thm:potter}
Let $h$ be a regularly varying function with index of variation $\rho\in\mathbb{R}$. For any $A>1, \delta>0$ there exists $X=X(A,\delta)$ such that, for all $x,y\geq X$,

$$\frac{h(y)}{h(x)} \leq A \max\left\{\left(\frac{y}{x}\right)^{\rho+\delta},\left(\frac{y}{x}\right)^{\rho-\delta}\right\}.$$
\end{proposition}

The following proposition is similar to Theorem 2.1 in \citet{Bingham2006}, which applies to convolutions of probability density functions. Here $h$ need not be a pdf.
\begin{proposition}
\label{thm:RVGauss}

Let $h:\mathbb R\to[0,\infty)$ be a locally integrable function such that $h(x)\sim x^\rho\ell(x)$ as $x\to\infty$, for some $\rho\in\mathbb R$ and some slowly varying function $\ell$. Assume also that $h(x)=O(1)$ as $x\to-\infty$. Then
 \begin{align}
 \frac
{\int_{-\infty}^\infty\phi(y-x)h(x)dx}{h(y)}\to 1
\end{align}
as $y\to\infty$, where $\phi$ is the pdf of a standard normal.
\end{proposition}

\begin{proof} The proof is similar to that of \citet[Theorem 2.1]{Bingham2006}. $h(x)=O(1)$ as $x\to-\infty$. Therefore, there is $X_0>0$ and $M>0$ such that $h(x)\leq M$ for all $x<-X_0$. For $y>0$,

{\small
\begin{align*}
\int_{-\infty}^\infty \frac{\phi(y-x)h(x)}{h(y)}dx &= \int_{-\infty}^{-X_0}\frac{\phi(y-x)h(x)}{h(y)}dx+ \int_{-X_0}^{y/2}\frac{\phi(y-x)h(x)}{h(y)}dx + \int_{y/2}^{\infty}\frac{\phi(y-x)h(x)}{h(y)}dx\\
&=\underbrace{\int_{-\infty}^{-X_0}\frac{\phi(y-x)h(x)}{h(y)}dx}_{A_3(y)}+ \underbrace{\int_{-X_0}^{y/2}\frac{\phi(y-x)h(x)}{h(y)}dx}_{A_2(y)} + \underbrace{\int_{-\infty}^{y/2}\frac{\phi(x)h(y-x)}{h(y)}dx}_{A_1(y)}
\end{align*}}

First consider $A_1(y)$. We apply Potter's theorem (\cref{thm:potter}) with $\delta=1+|\rho|$. There is $X_1>0$ such that, for all $u,v\geq X_1$,
$$
\frac{h(v)}{h(u)} \leq 2 \max\left\{\left(\frac{v}{u}\right)^{1+\max(2\rho,0)},\left(\frac{v}{u}\right)^{-1+\min(2\rho,0)}\right\}.
$$

 So for $y\geq2 X_1$ we have
$$0\leq x \leq \frac{y}{2} \Longrightarrow \frac{1}{2}\leq \frac{y-x}{y} \leq 1 \Longrightarrow \frac{h(y-x)}{h(y)}\leq 2\left(\frac{y-x}{y}\right)^{-1+\min(2\rho,0)} \leq 2^{2(1+|\rho|)}$$
and
$$ x \leq 0 \Longrightarrow 1\leq \frac{y-x}{y} \Longrightarrow \frac{h(y-x)}{h(y)}\leq 2\left(\frac{y-x}{y}\right)^{1+\max(2\rho,0)}\leq
2\left(1-\frac{x}{2X_1}\right)^{1+2|\rho|}.
$$
So for $y\geq2 X_1$ and $x\in\mathbb{R}$ we have
$$0\leq \frac{\phi(x)h(y-x)}{h(y)}\1{x\leq y/2} \leq \phi(x)\times\max(2^{2(1+|\rho|)},P(x))$$
where $P(x)=2\left(1-\frac{x}{2X_1}\right)^{1+2|\rho|}$ is a polynomial of degree $1+2|\rho|$ whose coefficients do not depend on $y$. Notice that $\phi\times P$ is integrable on $\mathbb{R}$ and, for any $x$,
$$
\lim_{y\to\infty}\frac{h(y-x)}{h(y)}= 1
$$
by \cref{thm:UCT}. Hence by the dominated convergence theorem we have
$$
\lim_{y\to \infty} A_1(y)=\int_{\mathbb R} \phi(x)dx=1.
$$

Consider now $A_2(y)$. For any $\epsilon>0$, we may find $X_2>X_1$ such that $\phi(x)\leq\epsilon \frac{h(x)}{(1+X_0+x)^{3(1+|\rho|)}}$ for all $x\geq  X_2$ and then for $y>2X_2$ we have

$$0\leq A_2(y) \leq \epsilon \int_{-X_0}^{y/2}\frac{h(x)}{h(y)}\frac{h(y-x)}{(1+X_0+y-x)^{3(1+|\rho|)}}dx \leq \epsilon \int_{-X_0}^{y/2}\frac{h(x)}{h(y)}\frac{h(y-x)}{(1+X_0+x)^{3(1+|\rho|)}}dx.$$

Indeed for $x<y/2$, $y>0$, and $c\geq 0$, $(c+y-x)^2-(c+x)^2=(2c+y)(y-2x)>0$ hence $0<(1+X_0+x)^{3(1+|\rho|)} < (1+X_0+y-x)^{3(1+|\rho|)}$. Using the same Potter's bound as for $A_1(y)$, we obtain that for any $y>X_2$

$$0\leq \int_{-X_0}^{y/2}\frac{h(y-x)h(x)}{h(y)(1+X_0+x)^{3(1+|\rho|)}}dx \leq \int_{-X_0}^\infty\frac{h(x)}{(1+X_0+x)^{3(1+|\rho|)}}\max(2^{2(1+|\rho|)},P(x))dx.$$

As $H(x)=\frac{h(x)}{(1+X_0+x)^{3(1+|\rho|)}}\times\max(2^{2(1+|\rho|)},P(x))=O(x^{-2})$ as $x\to\infty$, $H$ is integrable on $[-X_0,\infty)$ and $$\underset{y\rightarrow\infty}{\lim\sup }~A_2(y) =0.$$
Finally,
\begin{align}
A_3(y)=\int_{-\infty}^{-X_0}\frac{\phi(y-x)h(x)}{h(y)}dx\leq \frac{M}{h(y)}\int_{-\infty}^{-X_0} \phi(y-x)dx\leq \frac{M\phi(y+X_0)}{(y+X_0)h(y)}\to 0\text{ as }y\to\infty.
\end{align}
\end{proof}

\section{Experimental details}\label{sec:experimental_details}
\subsection{Implementation}

Code implementing our methods is written in Python and made available at \url{https://github.com/stefanocortinovis/robustcs/}.
All simulations were run locally on an Apple Silicon M4 Pro CPU with 24 GB of memory.

\subsection{Computation of eVCS}

In general, extended Ville confidence sequences cannot be expressed as intervals.
In particular, for a fixed $n$, evaluating an eVCS according to \cref{def:eVCS} requires constructing a grid over $\Theta$ and, for each grid element $\theta_0$, evaluating whether $\theta_0$ falls inside the region by checking if $m_n(y_{1:n})/f_{n,\theta_0}(y_{1:n}) \leq g_{\theta_0}^{-1}(\alpha)$.
Under a proper prior, the eVCS is guaranteed to be contained in the VCS (\cref{prop:BAestimatorinCS}), which is in turn an interval.
While this can be used to restrict the region over which the grid is constructed, fine gridding may still be computationally expensive and inconvenient in practical applications.

Alternatively, one can approximate an eVCS under a proper prior with an interval by means of common root-finding algorithms.
In particular, since the associated posterior mean is contained in the eVCS (\cref{prop:BAestimatorinCS}), one can use the latter together with the two VCS end points to construct two valid brackets for finding two roots, in $\theta_0$, of $m_n(y_{1:n})/f_{n,\theta_0}(y_{1:n}) = g_{\theta_0}^{-1}(\alpha)$.
Then, the interval having the two roots as end points is an approximation of the true eVCS, which matches it exactly when the latter is an interval, such as under the conditions of \cref{th:suffconditionprior}.

Unless stated otherwise, all empirical eVCSs reported in this work were found numerically to be intervals via gridding, and subsequently recomputed using the method described above, where Chandrupatla's bracketing algorithm \citep{Chandrupatla1997} was used for root-finding.
A tailored synthetic example of a mixture prior resulting in a disconnected eVCS is presented in \cref{sec:disconnected_cs}.

\section{Additional results}\label{sec:additional_results}
\subsection{Longer time horizon}
\begin{figure}
  \centering
  \includegraphics[width=\textwidth]{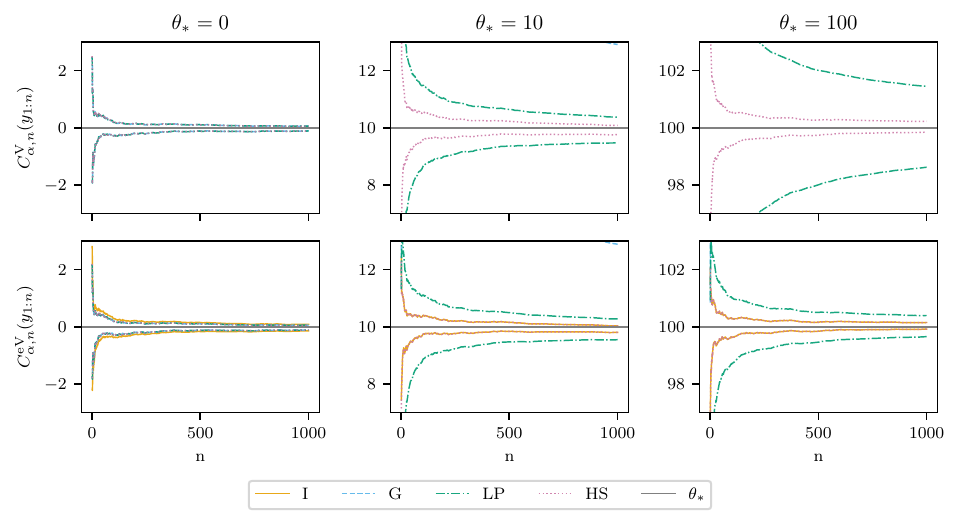}
  \caption{
    Same CSs as in Figure 1 run for $n$ up to 1000 instead of 100.
  }
  \label{fig:cs_intro_long}
\end{figure}

\Cref{fig:cs_intro} in the main text only considers $n$ up to 100 for better visibility of the differences between the methods.
To visualise the larger-sample behaviour of the CSs we consider, \cref{fig:cs_intro_long} replicates \cref{fig:cs_intro} with $n$ up to 1000.
In all cases, as $n$ increases, the CS eventually becomes more concentrated around the true $\thetatrue$, although differences between the methods, especially for $\thetatrue \in \{10, 100\}$, remain visible.

\subsection{Example of disconnected eVCS}\label{sec:disconnected_cs}

\begin{figure}[h!]
  \centering
  \includegraphics[width=\textwidth]{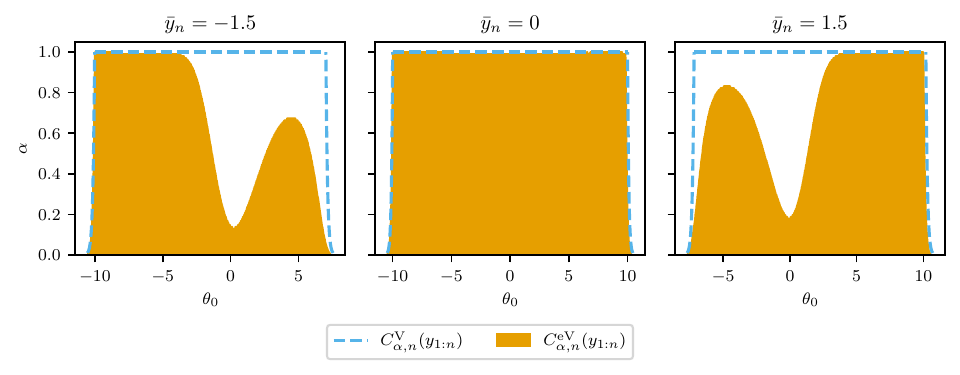}
  \caption{
    VCS and eVCS for the mean of a Gaussian with variance $\sigma^2=1$ under the asymmetric two-component Gaussian mixture prior $0.8 \times \mathcal{N}(-10, 10^{-4}) + 0.2 \times \mathcal{N}(10, 10^{-4})$, for $n = 1$, $\bar y_n \in \{-1.5, 0, 1.5\}$ and for various error levels $\alpha \in (0, 1)$.
    For $\bar y_n \in \{-1.5, 1.5\}$, the eVCS is clearly disconnected for $\alpha$ between approximately 0.2 and 0.6, with both mixture components contributing to the resulting CS.
  }
  \label{fig:cs_disconnected}
\end{figure}

While in all cases reported elsewhere in this work, the eVCSs were found to be intervals, this is not guaranteed in general, and eVCSs can be disconnected sets under particular priors.
One such example is given by the asymmetric two-component Gaussian mixture used in \cref{fig:cs_disconnected}, which, for some values of $\alpha$ and $\bar y_n$, clearly leads to a disconnected eVCS.
While a similar behaviour may arise for priors falling outside the assumptions of \cref{th:suffconditionprior}, we did not observe it in our experiments with heavier-tailed symmetric unimodal priors, and characterising the full class of priors leading to interval eVCSs is an open question.

\bibliographystyle{plainnat}
\bibliography{confidencesequences.bib}       

\end{document}